\documentclass[12pt]{article}
\usepackage{amsmath,amsfonts,amssymb,mathtools,latexsym}
\usepackage[nosort]{cite}
\usepackage{hyperref}
\hypersetup{hidelinks}
\usepackage{graphicx}
\usepackage{color}
\usepackage{epsfig}
\usepackage{psfrag}	
\usepackage{tikz}
\usetikzlibrary{calc}
\usepackage{slashed}
\usepackage{bbold}
\usepackage{tensor}
\usepackage{ulem}
\usepackage[font=small,labelfont=bf,width=1\textwidth]{caption}
\usepackage{comment}
\usepackage{dsfont}
\usepackage[all]{xy}
\usepackage{multirow}
\usepackage{float}
\usepackage{longtable}
\usepackage{pifont}
\usepackage{titlesec}
\usepackage{authblk}

\usepackage{tocloft}
\usepackage{soul,xcolor}
\setstcolor{red}
\usepackage{appendix}
\usepackage{etoolbox}

\definecolor{ZKgreen}{rgb}{0,0.6,0.3}

\tikzset{
  vtx/.style={
    circle,
    draw=blue,
    fill=blue,
    inner sep=1pt
  },
  wcirc/.style={
    circle,
    draw=white,
    fill=white,
    inner sep=2pt
  },
  bcirc/.style={
    circle,
    draw=black,
    fill=black,
    inner sep=1pt
  },
  dcirc/.style={
    circle,
    draw=blue,
    fill=blue,
    inner sep=1pt
  },
  rcirc/.style={
    circle,
    draw=red,
    fill=red,
    inner sep=1pt
  },
  phi/.style={
    thick
  },
  sigma/.style={
    thick,
    dashed
  },
  vl1/.style={
    thick,
    blue
  },
  vl2/.style={
    thick,
    dashed,
    blue
  },
  valign/.style={
    baseline={([yshift=-.55ex]current bounding box.center)}
  }
}

\newcommand\phiHphiHdiagA{
\begin{tikzpicture}[baseline]
  \draw[thick] (0, 0) -- (1.4, 0);
  \draw[dashed] (0.2, 0) to[out=90,in=90] (1.2, 0);
\end{tikzpicture} 
}

\newcommand\phiHphiHdiagB{
\begin{tikzpicture}[baseline]
  \draw[thick] (0, 0) -- (1.4, 0);
  \draw[dashed] (0.2, 0) to[out=90,in=90] (1.2, 0);
  \draw[dashed] (0.5, 0) -- (0.7, 0.25);
  \draw[dashed] (0.9, 0) -- (0.7, 0.25);
  \node at (0.7, 0.28) [bcirc] {};
  \node at (0.5, 0.00) [dcirc] {};
  \node at (0.9, 0.00) [dcirc] {};
\end{tikzpicture} 
}

\newcommand\threeptatep{
\begin{tikzpicture}[baseline]
  \draw[thick] (0, 0) -- (1.4, 0);
  \draw[dashed] (0.2, 0) to[out=90,in=90] (1.2, 0);
  \draw[dashed] (0.5, 0) -- (0.7, 0.25);
  \draw[dashed] (0.9, 0) -- (0.7, 0.25);
  \node at (0.7, 0.28) [bcirc] {};
  \node at (0.9, 0.00) [dcirc] {};
\end{tikzpicture}
}

\newcommand\connectedfourpt{
\begin{tikzpicture}[baseline]
  \draw[thick] (-0.4, 0) -- (1.4, 0);
  \draw[dashed] (-0.2, 0) -- (0.5, 0.7) -- (1.2, 0);
  \draw[dashed] ( 0.3, 0) -- (0.5, 0.7) -- (0.7, 0);
  \node at (0.5, 0.7) [bcirc] {};
\end{tikzpicture} 
}

\AtBeginEnvironment{subappendices}{%
  \addtocontents{toc}{\protect\addvspace{10pt}Appendices}
}
\makeatletter
\patchcmd{\@chapter}{\protect\numberline{\thechapter}#1}
{\@chapapp~\thechapter: #1}{}{}
\makeatother

\numberwithin{equation}{section}

\newcommand{\bea}{\begin{eqnarray}}
\newcommand{\eea}{\end{eqnarray}}
\newcommand{\beq}{\begin{equation}}
\newcommand{\eeq}{\end{equation}}
\newcommand{\bal}{\begin{equation}\begin{aligned}}
\newcommand{\eal}{\end{aligned} \end{equation}}

\title{\begin{flushright}
\end{flushright}
\vspace{2 cm}
Perturbative Defect CFT Data in Scalar Models and Consistency Conditions
\vspace{5mm}}
\author{Ziwen Kong%
\thanks{\href{mailto:zwn.kong@gmail.com}{zwn.kong@gmail.com}}\normalsize}
\affil{\it INFN, Sezione di Milano--Bicocca,\protect\\
Piazza della Scienza 3, 20126 Milano, Italy\vspace{4pt}}
\date{}

\begin{document}
\maketitle

\begin{abstract}
We compute defect CFT data through \(O(\epsilon)\) for scalar line defects in the
\(O(N)\), hypercubic, hypertetrahedral, \(O(m)\times O(n)\) biconical,
and \(MN\) models in \(d=4-\epsilon\). The \(O(N)\) theory provides a
benchmark; the datasets for the other four models are new. For the branches
considered, we determine spectra, operator mixing, two- and three-point
coefficients in a specified perturbative basis, and four-point functions
with identical or paired external
operators of integer dimension \(n\leq2\) at \(\epsilon=0\).
We propose a framework for verifying perturbative defect CFT data by
combining crossing symmetry, OPE reconstruction, and the applicable
broken-symmetry integral identities.
We also identify a new universal defect-direction sector whose one-loop
data follow from the defect fixed-point equation.
All applicable conditions are satisfied at the perturbative orders
considered. The model data are collected in the supplemental notebook,
\texttt{defect\_cft\_data.nb}.
\end{abstract}

\clearpage

\tableofcontents

\clearpage


\section{Introduction}
\label{sec:introduction}

Conformal line defects correspond to renormalization-group fixed points
of interactions localized on a line in a critical bulk theory. Even when the bulk
theory is fixed, different defect interactions can lead to distinct defect
fixed points, with different local spectra and correlation functions.
Scalar line defects offer a setting
in which perturbative calculations and bootstrap constraints can be studied
together, as illustrated by localized magnetic-field defects in critical
scalar theories~\cite{Allais:2014fqa,Cuomo:2021kfm,Gimenez-Grau:2022czc,Belton:2025hbu,Nishioka:2022qmj,Gimenez-Grau:2022ebb}.

We consider line defects generated by a linear coupling to bulk scalar
fields,
\begin{equation}
S_{\rm defect}
=h^I\int d\tau\,\phi_I(\tau,0).
\label{eq:defect-action}
\end{equation}
The bulk theories are studied near four dimensions using the
\(\epsilon\)-expansion~\cite{Wilson:1971dc}. Their quartic interactions
admit fixed points with different continuous and discrete internal
symmetries~\cite{Vicari:2007ma,Osborn:2017ucf,Rychkov:2018vya,Osborn:2020cnf}.
The bulk action is
\begin{equation}
S_{\rm bulk}
=
\int d^d x\,
\left[
\frac12
\partial^\mu\phi^I\partial_\mu\phi_I
+
\frac{1}{24}
\lambda^{IJKL}
\phi_I\phi_J\phi_K\phi_L
\right],
\qquad
d=4-\epsilon .
\label{eq:bulk-action}
\end{equation}
Here \(\phi_I\) are real scalar fields and \(\lambda^{IJKL}\) is the
quartic coupling tensor.

Our choice of bulk models and defect branches is based primarily on the
scalar examples studied in~\cite{Pannell:2023pwz}, where the analysis of
defect RG flows identifies fixed points and their stability in the five model classes considered here. Further results on
defect stability, including the hypertetrahedral model, are given
in~\cite{Pannell:2024hbu}. We take this classification as the starting
point for calculating local spectra and correlation functions on selected
defect branches. Comparing these data across bulk theories allows us to
distinguish a common sector from data sensitive to the bulk symmetry and
the chosen defect branch.

In this work we compute defect CFT data in five classes of
critical scalar theories in \(d=4-\epsilon\): the \(O(N)\), hypercubic,
hypertetrahedral, \(O(m)\times O(n)\) biconical, and \(MN\) models.
The datasets for the latter four models are new; the \(O(N)\) calculation
provides a benchmark against existing results.
This defect has been studied through
renormalization-group methods and conformal symmetry, including its local
spectrum and operator mixing~\cite{Cuomo:2021kfm,Nishioka:2022qmj}.
Analytic bootstrap and integral identities resolve further nearly degenerate
sectors~\cite{Belton:2025hbu}, while bulk two-point functions provide
complementary probes of the same defect~\cite{Gimenez-Grau:2022ebb}.

For the fixed-point branches included in our calculation, we determine
two-, three-, and four-point functions of operators with integer dimensions
\begin{equation}
n\leq2
\label{eq:intro-dimension-range}
\end{equation}
at \(\epsilon=0\). We denote their full scaling dimensions by \(\Delta\).
The four-point functions involve four identical operators or two pairs
of identical operators.
The operators include elementary scalar fields, their transverse
derivatives, and bilinears in representations of the preserved internal
symmetry. These include the displacement operator and, when continuous
internal symmetries are broken, tilt operators. Operators with identical preserved quantum numbers can mix, and
several sectors are degenerate at leading order. We give their operator
bases and mixing matrices together with the resulting scaling dimensions
and correlation functions. To evaluate the integral identities, we also compute the \(DDX\)
three-point coefficients and anomalous dimensions of the relevant
exchanged operators with \(n_X\leq3\).
The supplemental notebook collects the data for each model and branch.

Through \(O(\epsilon)\), we compute the following quantities
\begin{equation}
\left\{
\Delta_A,\,
C_{AB},\,
\lambda_{ABC},\,
\mathcal G_{ABCD}(\chi)
\right\}.
\label{eq:intro-cft-data}
\end{equation}
Here \(\Delta_A\) are scaling dimensions, \(C_{AB}\) are two-point
normalizations, \(\lambda_{ABC}\) are unit-normalized three-point
coefficients with the tensor structures factored out, and \(\mathcal G_{ABCD}(\chi)\) are reduced
four-point functions. In one dimension the four-point kinematics depend
on a single cross-ratio. This makes the relation between directly computed
correlators and the defect OPE particularly accessible, while the variety
of preserved symmetry groups supplies examples with nontrivial tensor
structures and operator mixing.

The size of this dataset motivates three complementary levels of
verification, each testing additional information in the computed CFT data.
Crossing symmetry provides the first level by checking the cross-ratio
dependence and tensor structures of each four-point function without
requiring a separately calculated spectrum. It establishes the internal
consistency of these functions under permutations of the external operators.

The OPE expansion adds a comparison with the operator dimensions and
OPE coefficients obtained independently from two- and three-point functions,
thereby connecting all three calculations. When exchanged operators have the same
leading-order dimension, their contributions cannot always be distinguished
at the perturbative order considered, so this test constrains weighted
sums of their CFT data. As the integer dimension of the exchanged operators increases, more
operators generally contribute to each degenerate sector, while the OPE
expansion of the same correlators at a given perturbative order probes
the same number of weighted sums. These sums do not, in general, determine
the individual operator dimensions and OPE coefficients, and the growing
number of exchanged operators makes this limitation more severe.

For correlators involving tilt or displacement operators, integral
identities extend this verification by supplying independent constraints
on further combinations of the spectrum and OPE coefficients. These identities follow from
broken continuous internal symmetries and transverse translations,
respectively, and relate integrated four-point functions to lower-point
data~\cite{Belton:2025hbu,Drukker:2022pxk,Sakkas:2024dvm,Gabai:2025zcs,Kong:2025sbk,Girault:2025kzt,Drukker:2025dfm,Belton:2025ief}.
They are nonperturbative and apply across models with the corresponding
broken symmetries. Their perturbative expansion probes combinations of CFT data beyond
those accessible from the ordinary OPE expansion at the orders considered.

The OPE and integral conditions constrain different weighted sums of the
same CFT data. For exchanged operators \(X_a\) with a common integer
dimension, the ordinary OPE expansion probes combinations such as
\begin{equation}
\sum_a\lambda_{ABa}\lambda_{CDa},
\qquad
\sum_a\lambda_{ABa}\lambda_{CDa}(\Delta_a-n_a),
\end{equation}
where \(\lambda_{ABa}\equiv\lambda_{ABX_a}\) and \(\Delta_a-n_a\)
is the anomalous dimension of \(X_a\). Integral identities instead weight the OPE products by integrated
conformal blocks. Analytic continuation of endpoint contributions gives
simple poles in the non-logarithmic kernels, producing terms schematically
of the form
\begin{equation}
\sum_a\frac{\lambda_{ABa}\lambda_{CDa}}
{\Delta_a-n_a},
\label{eq:intro-integral-simple-pole}
\end{equation}
while logarithmic kernels can contain double poles,
\begin{equation}
\sum_a\frac{\lambda_{ABa}\lambda_{CDa}}
{(\Delta_a-n_a)^2}.
\label{eq:intro-integral-double-pole}
\end{equation}
Equations~\eqref{eq:intro-integral-simple-pole} and
\eqref{eq:intro-integral-double-pole} schematically indicate the endpoint pole terms,
not the full integral identities; the exact denominators also depend on
the external dimensions and integration kernel.
Unlike the ordinary OPE expansion, the additional constraints from these
singular endpoint contributions involve exchanged operators within a
dimension range set by the endpoint convergence conditions. In the
\(D\times D\) channel, the relevant operators satisfy \(n_a\leq3\).

Together, crossing symmetry, OPE reconstruction, and the applicable
integral identities form our proposed framework for verifying the computed
observables in \eqref{eq:intro-cft-data}. The information they
test depends on the selected correlators, symmetry representations, and
perturbative order. The examples below illustrate their implementation
in sectors with mixing and broken symmetries, using the conventions of
the supplemental material.

Alongside these consistency conditions, we identify a new universal sector
whose CFT data can be predicted from the defect fixed-point equation.
The linear defect coupling selects a direction \(h_*^I\) in the space of
bulk scalar fields. Projecting the scalar fields and their transverse
derivatives along this direction gives a family of defect operators,
including the scalar \(\phi_0\) and the displacement \(D_\mu\).
We also retain the transverse-spin-two derivative operator
\(\mathcal D_{\mu\nu}\) as auxiliary endpoint data for the integral
identities.

The defect fixed-point equation makes the one-loop data in this sector
independent of the detailed bulk symmetry, apart from their dependence
on \(h_0\), the magnitude of the fixed-point defect coupling.
The value of \(h_0\) depends on the model and defect
branch, so universality here concerns the functional form rather than
identical numerical data. These predictions provide an additional test
of the model-specific calculations. The remaining sectors distinguish
the five models through their symmetry-breaking patterns, mixing blocks,
and inequivalent defect fixed points.

The calculations were carried out with symbolic computation and AI
assistance, using GPT 5.5, GPT 5.6 Sol, GPT 6 Sol, GPT 6.1 Sol, and
Opus 5.5.

The paper is organized as follows. Section~\ref{sec:consistency-conditions}
sets out the crossing, OPE, and integral conditions, including the
perturbative treatment of endpoint singularities.
Section~\ref{sec:universal-sector} presents the universal defect-direction
data and their consistency relations.
Section~\ref{sec:model-dependent-checks} describes the five models and
gives representative applications in sectors with mixing and broken
internal symmetries. Section~\ref{sec:discussion} discusses the information
constrained by the different conditions and the limits imposed by
perturbative order.


\section{Consistency conditions for perturbative defect CFT data}
\label{sec:consistency-conditions}

We begin with bare defect operators \(\mathcal O_A^{\rm bare}\) and
define the renormalized operators by
\begin{equation}
\mathcal O_A^{\rm bare}
=
Z_A{}^B \mathcal O_B ,
\label{eq:general-operator-renormalization}
\end{equation}
where \(Z_A{}^B\) is a matrix when operators with the same quantum numbers
mix.  Resolving this mixing gives operators of definite scaling dimension
\(\Delta_A\).  Denoting separations along the defect by
\(\tau_{ij}\equiv\tau_i-\tau_j\), we choose a basis in which the two-point
functions are diagonal,
\begin{equation}
\left\langle
\mathcal O_A(\tau_1)\mathcal O_B(\tau_2)
\right\rangle
=
\delta_{AB}
\frac{C_A}{|\tau_{12}|^{2\Delta_A}} .
\label{eq:general-two-point}
\end{equation}
Thus the two-point calculation determines, to the perturbative order under
consideration, the renormalization factors, scaling dimensions, and
coefficients \(C_A\).  The latter depend on the normalization convention for
generic defect primaries.  The displacement and tilt operators are
exceptions, since their normalization is fixed by the corresponding Ward
identities~\cite{Billo:2016cpy,Drukker:2022pxk}.

It is convenient to introduce unit-normalized operators
\begin{equation}
\widehat{\mathcal O}_A
\equiv
\frac{\mathcal O_A}{\sqrt{C_A}},
\label{eq:unit-normalized-operators}
\end{equation}
for which
\begin{equation}
\left\langle
\widehat{\mathcal O}_A(\tau_1)
\widehat{\mathcal O}_B(\tau_2)
\right\rangle
=
\frac{\delta_{AB}}
{|\tau_{12}|^{2\Delta_A}} .
\label{eq:unit-normalized-two-point}
\end{equation}

For these unit-normalized operators, the three-point functions take the form
\begin{equation}
\left\langle
\widehat{\mathcal O}_A(\tau_1)
\widehat{\mathcal O}_B(\tau_2)
\widehat{\mathcal O}_C(\tau_3)
\right\rangle
=
T_{ABC}\,
\frac{\lambda_{ABC}}
{|\tau_{12}|^{\Delta_A+\Delta_B-\Delta_C}
|\tau_{13}|^{\Delta_A+\Delta_C-\Delta_B}
|\tau_{23}|^{\Delta_B+\Delta_C-\Delta_A}},
\label{eq:general-three-point}
\end{equation}
where \(T_{ABC}\) denotes the invariant internal and transverse-spin tensor
structure, and \(\lambda_{ABC}\) is the scalar OPE coefficient.

We use this unit-normalized convention whenever writing
conformal-block decompositions below.

Four-point functions bring these ingredients together.  They depend on the
cross-ratio
\begin{equation}
\chi
=
\frac{\tau_{12}\tau_{34}}
{\tau_{13}\tau_{24}} .
\label{eq:cross-ratio}
\end{equation}
For four unit-normalized defect primaries of dimensions \(\Delta_i\), we
write
\begin{equation}
\left\langle
\widehat{\mathcal O}_1(\tau_1)
\widehat{\mathcal O}_2(\tau_2)
\widehat{\mathcal O}_3(\tau_3)
\widehat{\mathcal O}_4(\tau_4)
\right\rangle
=
\frac{1}
{|\tau_{12}|^{\Delta_1+\Delta_2}
|\tau_{34}|^{\Delta_3+\Delta_4}}
\left|
\frac{\tau_{24}}{\tau_{14}}
\right|^{\Delta_1-\Delta_2}
\left|
\frac{\tau_{14}}{\tau_{13}}
\right|^{\Delta_3-\Delta_4}
\sum_R T_R\,\mathcal G_R(\chi),
\label{eq:general-four-point}
\end{equation}
where \(T_R\) form a basis of invariant tensor structures.

After projection onto a fixed tensor structure and exchanged
representation, the \(12\to34\) channel takes schematically the form
\begin{equation}
\mathcal G_{1234}(\chi)
=
\sum_X
\lambda_{12X}\lambda_{34X}\,
g_{\Delta_X}^{\Delta_{12},\Delta_{34}}(\chi),
\label{eq:general-ope-decomposition}
\end{equation}
where \(\Delta_{ij}\equiv\Delta_i-\Delta_j\), and
\begin{equation}
g_{\Delta}^{\Delta_{12},\Delta_{34}}(\chi)
=
\chi^\Delta\,
{}_2F_1
\left(
\Delta-\Delta_{12},
\Delta+\Delta_{34};
2\Delta;
\chi
\right).
\label{eq:one-dimensional-block}
\end{equation}
The sum is restricted by the relevant symmetry representations and selection
rules.

The renormalized four-point functions are obtained using the same
renormalization factors fixed by the two-point calculation.  Independently,
the spectrum and scalar OPE coefficients extracted from the two- and
three-point functions predict their conformal-block decomposition.
Their agreement relates the independently computed two-, three-, and
four-point data.

\subsection{Crossing symmetry}
\label{subsec:crossing}

Crossing symmetry provides an internal check of the four-point functions.
A permutation of the external operators acts simultaneously on the
cross-ratio, conformal prefactor, and invariant tensor structures.  The
resulting expression must agree with the original correlator whenever the
permutation leaves the external operator content unchanged.

In this work, we restrict the four-point calculation to correlators with
either four identical external operators or two pairs of identical external
operators.  In both cases, exchanging two identical operators gives
\begin{equation}
\chi
\longrightarrow
\frac{\chi}{\chi-1},
\label{eq:crossing-map-pair}
\end{equation}
together with the corresponding transformation of the conformal prefactor
and tensor structures.  When all four external operators are identical,
there is an additional crossing relation associated with
\begin{equation}
\chi
\longrightarrow
1-\chi .
\label{eq:crossing-map-identical}
\end{equation}

We impose these crossing relations component by component in the tensor
basis and order by order in \(\epsilon\).  In the complete dataset described
in the supplemental material, every computed four-point function satisfies
the applicable crossing relations to the perturbative order considered.

\subsection{OPE expansion and consistency}
\label{subsec:ope-consistency}

Beyond crossing symmetry, we compare the computed four-point functions
with their OPE expansions, using the spectrum and OPE coefficients obtained
independently from the two- and three-point calculations.

In sectors degenerate at \(\epsilon=0\), one-loop mixing fixes the
leading eigenvectors. Their \(O(\epsilon)\) rotations can depend on the
\(O(\epsilon^2)\) mixing matrix. Unless these rotations are fixed,
individual \(O(\epsilon)\) OPE corrections with nonzero tree-level
coefficients refer to the specified one-loop eigenbasis, rather than
fully determined coefficients of the exact scaling eigenoperators.
The comparisons below test basis-independent combinations.
For OPE coefficients starting at \(O(\epsilon)\), these rotations first
change the \(O(\epsilon^2)\) term.

For the correlators considered in this work, the external
operators are identical in pairs in the \(12\to34\) channel, so that
\begin{equation}
\Delta_{12}=\Delta_{34}=0 .
\end{equation}

We write the scaling dimension of an exchanged operator \(X\) as
\begin{equation}
\Delta_X
=
n_X
+
\epsilon\,\gamma_X
+
O(\epsilon^2),
\label{eq:dimension-expansion}
\end{equation}
where \(n_X\) is its integer dimension at \(\epsilon=0\).  The corresponding
conformal block expands as
\begin{equation}
g_{\Delta_X}(\chi)
=
g_{n_X}(\chi)
+
\epsilon\,\gamma_X
\left.
\partial_\Delta g_\Delta(\chi)
\right|_{\Delta=n_X}
+
O(\epsilon^2).
\label{eq:block-expansion}
\end{equation}
We similarly expand
\begin{equation}
\lambda_{12X}\lambda_{34X}
=
\bigl(\lambda_{12X}\lambda_{34X}\bigr)^{(0)}
+
\epsilon\,
\bigl(\lambda_{12X}\lambda_{34X}\bigr)^{(1)}
+
O(\epsilon^2).
\label{eq:ope-product-expansion}
\end{equation}
The coefficients of this product are
\begin{align}
\bigl(\lambda_{12X}\lambda_{34X}\bigr)^{(0)}
&=\lambda_{12X}^{(0)}\lambda_{34X}^{(0)},\\
\bigl(\lambda_{12X}\lambda_{34X}\bigr)^{(1)}
&=\lambda_{12X}^{(0)}\lambda_{34X}^{(1)}
+\lambda_{12X}^{(1)}\lambda_{34X}^{(0)}.
\end{align}
The contribution of \(X\) is therefore
\begin{align}
\lambda_{12X}\lambda_{34X}\,
g_{\Delta_X}(\chi)
={}&
\bigl(\lambda_{12X}\lambda_{34X}\bigr)^{(0)}
g_{n_X}(\chi)
 +
\epsilon\,
\bigl(\lambda_{12X}\lambda_{34X}\bigr)^{(1)}
g_{n_X}(\chi)
\nonumber\\
&+
\epsilon\,\gamma_X
\bigl(\lambda_{12X}\lambda_{34X}\bigr)^{(0)}
\left.
\partial_\Delta g_\Delta(\chi)
\right|_{\Delta=n_X}
+
O(\epsilon^2).
\label{eq:perturbative-block-contribution}
\end{align}
Thus the logarithmic terms in the small-\(\chi\) expansion probe anomalous
dimensions weighted by leading OPE products.

The OPE sums over all exchanged primaries allowed by the external
operators and the selection rules. At each integer dimension, we collect
the contributions to the same tensor channel. If \(X_a\) denote the
operators in such a group, their contribution involves combinations
such as
\begin{equation}
\sum_a
\bigl(\lambda_{12X_a}\lambda_{34X_a}\bigr)^{(0)},
\qquad
\sum_a
\bigl(\lambda_{12X_a}\lambda_{34X_a}\bigr)^{(0)}
\gamma_a .
\label{eq:degenerate-spectral-sums}
\end{equation}
For a single exchanged primary these sums contain one term; when several
primaries share the same leading dimension, they combine their contributions.

To compare with the computed four-point function, we expand both sides
of the OPE decomposition at small \(\chi\) and match the coefficients
of powers of \(\chi\) and \(\log\chi\), order by order in
\(\epsilon\), in each tensor channel. Contributions of lower-dimensional
primaries and their descendants are included when identifying the
coefficients associated with a given leading dimension. This comparison tests the sums above and their perturbative corrections,
rather than the contribution of each exchanged operator separately.

For a fixed set of external operators and a fixed perturbative order,
the number of independent combinations resolving a given degenerate sector
is limited. The number of exchanged operators generally grows with their
classical dimension, so these combinations need not determine each
individual dimension or OPE coefficient. In this work, the ordinary OPE comparison is performed through
\(\chi^2\), using the three-point data for exchanged primaries with
\(n_X\leq2\). The additional \(DDX\) data with \(n_X\leq3\) are used
for the integral identities, rather than to extend this comparison.
The expansion is
\begin{align}
\mathcal G(\chi)=\mathcal G_{\mathbb 1}
&+\chi\sum_{n_a=1}\left[\bigl(\lambda_{12X_a}\lambda_{34X_a}\bigr)^{(0)}
+\epsilon\bigl(\lambda_{12X_a}\lambda_{34X_a}\bigr)^{(1)}\right]
\nonumber\\
&+\epsilon\chi\log\chi\sum_{n_a=1}
\bigl(\lambda_{12X_a}\lambda_{34X_a}\bigr)^{(0)}\gamma_a
\nonumber\\
&+\chi^2\left\{
\sum_{n_a=2}\left[\bigl(\lambda_{12X_a}\lambda_{34X_a}\bigr)^{(0)}
+\epsilon\bigl(\lambda_{12X_a}\lambda_{34X_a}\bigr)^{(1)}\right]
\right.
\nonumber\\
&\qquad\left. +\frac12\sum_{n_a=1}\left[\bigl(\lambda_{12X_a}\lambda_{34X_a}\bigr)^{(0)}
+\epsilon\bigl(\lambda_{12X_a}\lambda_{34X_a}\bigr)^{(1)}
+\epsilon\bigl(\lambda_{12X_a}\lambda_{34X_a}\bigr)^{(0)}\gamma_a\right]
\right\}
\nonumber\\
&+\epsilon\chi^2\log\chi\left[
\sum_{n_a=2}\bigl(\lambda_{12X_a}\lambda_{34X_a}\bigr)^{(0)}\gamma_a
+\frac12\sum_{n_a=1}\bigl(\lambda_{12X_a}\lambda_{34X_a}\bigr)^{(0)}\gamma_a
\right]
+O(\chi^3,\epsilon^2).
\end{align}
Here \(\mathcal G_{\mathbb 1}\) is the identity contribution in the
chosen tensor channel, and the factors of \(1/2\) arise from the first
descendant of the \(n_a=1\) block. When all four external operators
are identical, the OPE products reduce to \(\lambda_a^2\), with
\((\lambda_a^2)^{(1)}=2\lambda_a^{(0)}\lambda_a^{(1)}\).
The \(\chi\)
and \(\chi\log\chi\) terms test the \(n_a=1\) data; after accounting
for their descendant contributions, the \(\chi^2\) and
\(\chi^2\log\chi\) terms test the \(n_a=2\) data. If both OPE
coefficients start at \(O(\epsilon)\), their product contributes only
at \(O(\epsilon^2)\) and is not tested by this comparison.
The coefficients through \(\chi^2\) agree with the independently
computed data at the orders where they contribute. Examples are given
in sections~\ref{sec:universal-sector} and
\ref{sec:model-dependent-checks}.

\subsection{Integral identities}
\label{subsec:integral-identities}

The final class of consistency conditions originates from symmetries broken
by the defect.  A line defect breaks transverse translations and may also
break part of the internal symmetry of the bulk theory.  The corresponding
broken symmetries are represented on the defect by the displacement and tilt
operators, respectively. The associated integral identities are
nonperturbative relations applicable across models with the relevant broken
symmetries~\cite{Drukker:2022pxk,Gabai:2025zcs,Kong:2025sbk,Girault:2025kzt,Drukker:2025dfm,Belton:2025ief}.

For the line defects considered here, the relevant identities involve
four-point functions with two displacement or tilt insertions and two
additional defect operators.  Their evaluation therefore combines
separated-point four-point functions with endpoint data determined by local
operator dimensions and OPE coefficients.  They provide consistency
conditions involving independently computed two-, three-, and four-point
data.

We first discuss the identities associated with broken transverse
translations and then those associated with broken continuous internal
symmetries.
For applications of integral constraints to extracting perturbative
defect data, see~\cite{Belton:2025hbu,Sakkas:2024dvm,Bliard:2023zpe}.

\subsubsection{Displacement identities}

Integral constraints from nonlinearly realized ambient conformal symmetry
are developed in~\cite{Gabai:2025zcs,Drukker:2025dfm}; the displacement
normalization follows the defect Ward identity~\cite{Billo:2016cpy}.
\label{subsubsec:displacement-identities}

The displacement operator is defined by
\begin{equation}
\partial^\mu T_{\mu i}(x)
=
\delta^{(d-1)}(x_\perp)\,D_i(\tau),
\label{eq:displacement-ward-integral}
\end{equation}
where \(i=1,\ldots,d-1\).  For a conformal line defect it has protected
dimension \(\Delta_D=2\), and
\begin{equation}
\left\langle
D_i(\tau)D_j(0)
\right\rangle
=
\frac{C_D\,\delta_{ij}}{|\tau|^4}.
\label{eq:displacement-two-point-integral}
\end{equation}

Let \(\mathcal O_\alpha\) be a defect primary of dimension
\(\Delta_{\mathcal O}\), normalized by
\begin{equation}
\left\langle
\mathcal O_\alpha(0)\mathcal O_\beta(\infty)
\right\rangle
=
C_{\mathcal O}\,\delta_{\alpha\beta}.
\label{eq:O-two-point-integral}
\end{equation}
We define
\begin{align}
\left\langle
D_i(1)D_j(\tau)
\mathcal O_\alpha(0)\mathcal O_\beta(\infty)
\right\rangle_c
\equiv{}&
\left\langle
D_i(1)D_j(\tau)
\mathcal O_\alpha(0)\mathcal O_\beta(\infty)
\right\rangle
\nonumber\\
&-
\left\langle D_i(1)D_j(\tau)\right\rangle
\left\langle
\mathcal O_\alpha(0)\mathcal O_\beta(\infty)
\right\rangle
\nonumber\\
&-
\left\langle D_i(1)\mathcal O_\alpha(0)\right\rangle
\left\langle D_j(\tau)\mathcal O_\beta(\infty)\right\rangle
\nonumber\\
&-
\left\langle D_i(1)\mathcal O_\beta(\infty)\right\rangle
\left\langle D_j(\tau)\mathcal O_\alpha(0)\right\rangle .
\label{eq:connected-DDOO}
\end{align}

The non-log identities are
\begin{equation}
\int_{-\infty}^{\infty}d\tau\,
\tau^n
\left\langle
D_i(1)D_j(\tau)
\mathcal O_\alpha(0)\mathcal O_\beta(\infty)
\right\rangle_c
=
0,
\qquad
n=0,1,2,
\label{eq:displacement-nonlog-identities}
\end{equation}
while the log identities are
\begin{align}
\int_{-\infty}^{\infty}d\tau\,
\log|\tau|\,
\left\langle
D_i(1)D_j(\tau)
\mathcal O_\alpha(0)\mathcal O_\beta(\infty)
\right\rangle_c
={}&
\delta_{ij}\Delta_{\mathcal O}
C_{\mathcal O}\delta_{\alpha\beta}
-
\left\langle
[M_{ij},\mathcal O_\alpha](0)
\mathcal O_\beta(\infty)
\right\rangle ,
\nonumber\\
\int_{-\infty}^{\infty}d\tau\,
\tau\log|\tau|\,
\left\langle
D_i(1)D_j(\tau)
\mathcal O_\alpha(0)\mathcal O_\beta(\infty)
\right\rangle_c
={}&
\frac12
\left\langle
[M_{ij},\mathcal O_\alpha](0)
\mathcal O_\beta(\infty)
\right\rangle ,
\nonumber\\
\int_{-\infty}^{\infty}d\tau\,
\tau^2\log|\tau|\,
\left\langle
D_i(1)D_j(\tau)
\mathcal O_\alpha(0)\mathcal O_\beta(\infty)
\right\rangle_c
={}&
-\delta_{ij}\Delta_{\mathcal O}
C_{\mathcal O}\delta_{\alpha\beta}
-
\left\langle
[M_{ij},\mathcal O_\alpha](0)
\mathcal O_\beta(\infty)
\right\rangle .
\label{eq:displacement-log-identities}
\end{align}
Here \(M_{ij}\) generates rotations in the transverse \(ij\)-plane,
and \([M_{ij},\mathcal O_\alpha]\) denotes its action on the
transverse-spin representation of \(\mathcal O_\alpha\). The indices
\(i,j\) are transverse spatial indices. For a transverse
scalar, the commutator term vanishes.

For perturbative calculations, we split the real line into the three
ordering regions and map each to \(0<\tau<1\).  The non-log identities become
\begin{align}
0
={}&
\int_0^1 d\tau\,
(-1)^n\tau^n(1-\tau)^{2-n}
\left\langle
D_i(1)D_j(0)
\mathcal O_\alpha(\tau)\mathcal O_\beta(\infty)
\right\rangle_c
\nonumber\\
&+
\int_0^1 d\tau\,
\tau^n
\left\langle
D_i(1)D_j(\tau)
\mathcal O_\alpha(0)\mathcal O_\beta(\infty)
\right\rangle_c
\nonumber\\
&+
\int_0^1 d\tau\,
\tau^{2-n}
\left\langle
D_i(\tau)D_j(1)
\mathcal O_\alpha(0)\mathcal O_\beta(\infty)
\right\rangle_c ,
\qquad
n=0,1,2.
\label{eq:displacement-nonlog-01}
\end{align}
The left-hand sides of \eqref{eq:displacement-log-identities}, for
\(n=0,1,2\), become the following expression under the same mapping;
their right-hand sides are unchanged.
\begin{align}
&\int_0^1 d\tau\,
(-1)^n\tau^n(1-\tau)^{2-n}
\bigl[\log\tau-\log(1-\tau)\bigr]
\left\langle
D_i(1)D_j(0)
\mathcal O_\alpha(\tau)\mathcal O_\beta(\infty)
\right\rangle_c
\nonumber\\
&+
\int_0^1 d\tau\,
\tau^n\log\tau
\left\langle
D_i(1)D_j(\tau)
\mathcal O_\alpha(0)\mathcal O_\beta(\infty)
\right\rangle_c
\nonumber\\
&-
\int_0^1 d\tau\,
\tau^{2-n}\log\tau
\left\langle
D_i(\tau)D_j(1)
\mathcal O_\alpha(0)\mathcal O_\beta(\infty)
\right\rangle_c .
\label{eq:displacement-log-01}
\end{align}

The endpoint limits are controlled by the \(D\times D\) and
\(D\times\mathcal O\) OPEs.  For power counting it is useful to work with
unit-normalized primaries in the endpoint expansion,
\begin{align}
\widehat D_i(\tau_1)\widehat D_j(\tau_2)
&\sim
T_{ij;X}\,
\lambda_{DDX}\,
|\tau_{12}|^{\Delta_X-4}
\widehat X(\tau_2)+\cdots ,
\label{eq:DD-endpoint-ope}
\\
\widehat D_i(\tau_1)\widehat{\mathcal O}_\alpha(\tau_2)
&\sim
T_{i\alpha;X}\,
\lambda_{D\mathcal O X}\,
|\tau_{12}|^{\Delta_X-\Delta_{\mathcal O}-2}
\widehat X(\tau_2)+\cdots .
\label{eq:DO-endpoint-ope}
\end{align}
The physical normalizations \(C_D\) and \(C_{\mathcal O}\) are restored in
the full correlators and in the Ward-identity right-hand sides.

Let \(k\) denote the additional power of the OPE separation supplied by the
mapped measure.  A primary \(X\) in the \(D\times\mathcal O\) OPE behaves as
\begin{equation}
|\tau_{12}|^{\Delta_X-\Delta_{\mathcal O}-2+k}.
\label{eq:DO-endpoint-power}
\end{equation}
A contribution can require a distributional treatment only if
\begin{equation}
n_X+k\leq n_{\mathcal O}+1.
\label{eq:DO-singular-condition}
\end{equation}
For \(D\times D\),
\begin{equation}
n_X+k\leq3.
\label{eq:DD-singular-condition}
\end{equation}

For every contribution selected by this criterion, let
\(f_{\rm sing}(\tau)\) denote the corresponding singular OPE contribution.
We write
\begin{equation}
\int_0^1 d\tau\,f(\tau)
=
\int_0^1 d\tau\,
\bigl[f(\tau)-f_{\rm sing}(\tau)\bigr]
+
\int_0^1 d\tau\,f_{\rm sing}(\tau).
\label{eq:subtraction-addback}
\end{equation}
The first integral is regular and may be expanded in \(\epsilon\) before
integration.  The second is evaluated by analytic continuation while keeping
the full \(\epsilon\)-dependent dimensions unexpanded:
\begin{equation}
\int_0^1 d\tau\,
\tau^{\Delta_X-\Delta_{\mathcal O}-2+k}
=
\frac{1}
{\Delta_X-\Delta_{\mathcal O}-1+k}.
\label{eq:DO-meromorphic-power}
\end{equation}
For logarithmic weights,
\begin{align}
\int_0^1 d\tau\,
\tau^{\Delta_X-\Delta_{\mathcal O}-2+k}\log\tau
&=
-\frac{1}
{\bigl(\Delta_X-\Delta_{\mathcal O}-1+k\bigr)^2},
\label{eq:DO-meromorphic-log-same}
\\
\int_0^1 d\tau\,
\tau^{\Delta_X-\Delta_{\mathcal O}-2+k}\log(1-\tau)
&=
\frac{
\psi(1)-\psi(\Delta_X-\Delta_{\mathcal O}+k)
}{
\Delta_X-\Delta_{\mathcal O}-1+k
}.
\label{eq:DO-meromorphic-log-opposite}
\end{align}

Of particular importance is the borderline case
\begin{equation}
n_X-n_{\mathcal O}-1+k=0,
\label{eq:borderline-endpoint}
\end{equation}
for which
\begin{equation}
\Delta_X-\Delta_{\mathcal O}-1+k
=
\epsilon(\gamma_X-\gamma_{\mathcal O})
+
O(\epsilon^2).
\label{eq:borderline-endpoint-exponent}
\end{equation}
For the borderline contributions encountered below, both scalar OPE
coefficients start at \(O(\epsilon)\),
\begin{equation}
\lambda_{D\mathcal O X}\lambda_{\cdots X}
=
O(\epsilon^2).
\label{eq:borderline-ope-order}
\end{equation}
The non-log add-back is therefore enhanced by one inverse power of
\(\epsilon\),
\begin{equation}
\frac{
\lambda_{D\mathcal O X}\lambda_{\cdots X}
}{
\Delta_X-\Delta_{\mathcal O}-1+k
}
\sim
\frac{
\lambda_{D\mathcal O X}\lambda_{\cdots X}
}{
\epsilon(\gamma_X-\gamma_{\mathcal O})
}
=
O(\epsilon),
\label{eq:borderline-nonlog-scaling}
\end{equation}
whereas the same-endpoint logarithmic contribution is enhanced by two,
\begin{equation}
-\frac{
\lambda_{D\mathcal O X}\lambda_{\cdots X}
}{
\bigl(\Delta_X-\Delta_{\mathcal O}-1+k\bigr)^2
}
\sim
-\frac{
\lambda_{D\mathcal O X}\lambda_{\cdots X}
}{
\epsilon^2(\gamma_X-\gamma_{\mathcal O})^2
}
=
O(\epsilon^0).
\label{eq:borderline-log-scaling}
\end{equation}
For the opposite-endpoint logarithm the apparent singularity cancels,
\begin{equation}
\frac{
\psi(1)-\psi(\Delta_X-\Delta_{\mathcal O}+k)
}{
\Delta_X-\Delta_{\mathcal O}-1+k
}
=
-\psi'(1)+O(\epsilon).
\label{eq:borderline-opposite-log-scaling}
\end{equation}

These endpoint enhancements must be included when determining which CFT
data contribute at a given perturbative order.

\subsubsection{Tilt identities}
\label{subsubsec:tilt-identities}

When the defect breaks a continuous internal symmetry, the broken currents
define tilt operators through
\begin{equation}
\partial^\mu J_\mu^A(x)
=
\delta^{(d-1)}(x_\perp)\,
P^A{}_i\,t_i(\tau).
\label{eq:tilt-ward}
\end{equation}
For a conformal line defect, \(\Delta_t=1\), and
\begin{equation}
\left\langle
t_i(\tau)t_j(0)
\right\rangle
=
\frac{C_t\,\delta_{ij}}{|\tau|^2}.
\label{eq:tilt-two-point}
\end{equation}

The connected correlator obeys
\begin{equation}
\int_{-\infty}^{\infty}d\tau\,
\left\langle
t_i(1)t_j(\tau)
\mathcal O_\alpha(0)\mathcal O_\beta(\infty)
\right\rangle_c
=
0,
\label{eq:tilt-nonlog-identity}
\end{equation}
and
\begin{equation}
\int_{-\infty}^{\infty}d\tau\,
\log|\tau|\,
\left\langle
t_i(1)t_j(\tau)
\mathcal O_\alpha(0)\mathcal O_\beta(\infty)
\right\rangle_c
=
\frac12
\left\langle
[[Q_i,Q_j],\mathcal O_\alpha](0)
\mathcal O_\beta(\infty)
\right\rangle .
\label{eq:tilt-log-identity}
\end{equation}

Mapping to \(0<\tau<1\), the non-log identity becomes
\begin{align}
0
={}&
\int_0^1d\tau\,
\left\langle
t_i(1)t_j(0)
\mathcal O_\alpha(\tau)\mathcal O_\beta(\infty)
\right\rangle_c
\nonumber\\
&+
\int_0^1d\tau\,
\left\langle
t_i(1)t_j(\tau)
\mathcal O_\alpha(0)\mathcal O_\beta(\infty)
\right\rangle_c
\nonumber\\
&+
\int_0^1d\tau\,
\left\langle
t_i(\tau)t_j(1)
\mathcal O_\alpha(0)\mathcal O_\beta(\infty)
\right\rangle_c ,
\label{eq:tilt-nonlog-01}
\end{align}
while the log identity gives
\begin{align}
&\int_0^1d\tau\,
\bigl[\log\tau-\log(1-\tau)\bigr]
\left\langle
t_i(1)t_j(0)
\mathcal O_\alpha(\tau)\mathcal O_\beta(\infty)
\right\rangle_c
\nonumber\\
&+
\int_0^1d\tau\,
\log\tau
\left\langle
t_i(1)t_j(\tau)
\mathcal O_\alpha(0)\mathcal O_\beta(\infty)
\right\rangle_c
\nonumber\\
&-
\int_0^1d\tau\,
\log\tau
\left\langle
t_i(\tau)t_j(1)
\mathcal O_\alpha(0)\mathcal O_\beta(\infty)
\right\rangle_c
\nonumber\\
&=\frac12
\left\langle
[[Q_i,Q_j],\mathcal O_\alpha](0)
\mathcal O_\beta(\infty)
\right\rangle .
\label{eq:tilt-log-01}
\end{align}

Using unit-normalized operators for the endpoint OPE coefficients,
\begin{align}
\widehat t_i(\tau_1)\widehat t_j(\tau_2)
&\sim
T_{ij;X}\,
\lambda_{ttX}\,
|\tau_{12}|^{\Delta_X-2}
\widehat X(\tau_2)+\cdots ,
\label{eq:tt-endpoint-ope}
\\
\widehat t_i(\tau_1)\widehat{\mathcal O}_\alpha(\tau_2)
&\sim
T_{i\alpha;X}\,
\lambda_{t\mathcal O X}\,
|\tau_{12}|^{\Delta_X-\Delta_{\mathcal O}-1}
\widehat X(\tau_2)+\cdots .
\label{eq:tO-endpoint-ope}
\end{align}
Since the mapped tilt identities contain no additional non-logarithmic
endpoint powers, the potentially singular primary contributions satisfy
\begin{equation}
n_X\leq n_{\mathcal O}
\qquad
\text{in the }t\times\mathcal O\text{ OPE},
\label{eq:tO-singular-condition}
\end{equation}
and
\begin{equation}
n_X\leq1
\qquad
\text{in the }t\times t\text{ OPE}.
\label{eq:tt-singular-condition}
\end{equation}
In particular, the defect-direction scalar \(\phi_0\), with
\(n_{\phi_0}=1\), lies at the borderline in the \(t\times t\) channel.

All selected contributions are treated with the same subtraction and
analytic-continuation prescription as in the displacement sector.


\section{The universal defect-direction sector}
\label{sec:universal-sector}

We now specialize to scalar line defects of the form
\eqref{eq:defect-action} in the bulk theory defined by
\eqref{eq:bulk-action}.
The tensor coupling \(\lambda^{IJKL}\) appears directly in the bulk
action and includes the physical normalization. Model-specific scalar
couplings use this same convention.

At leading order,
\begin{equation}
\beta^I
=
-\frac{\epsilon}{2}h^I
+
\frac{1}{96\pi^2}
\lambda^{IJKL}h_Jh_Kh_L
+
O(\lambda^2).
\label{eq:defect-beta}
\end{equation}
A conformal line defect is obtained at
\begin{equation}
-\frac{\epsilon}{2}h_*^I
+
\frac{1}{96\pi^2}
\lambda_*^{IJKL}
h_{*,J}h_{*,K}h_{*,L}
=
0.
\label{eq:defect-fixed-point}
\end{equation}

The defect fixed point singles out a distinguished direction \(h_*^I\).
The corresponding projected family begins with
\begin{equation}
h_*^I\phi_I,
\qquad
h_*^I\partial_\mu\phi_I,
\qquad
h_*^I\partial_\mu\partial_\nu\phi_I,
\qquad
\ldots .
\label{eq:defect-direction-family}
\end{equation}
These operators are singlets under the internal symmetry preserved by the
defect.

Define
\begin{equation}
h_0^2
\equiv
h_*^Ih_{*,I},
\qquad
\Lambda_h
\equiv
\lambda_*^{IJKL}
h_{*,I}h_{*,J}h_{*,K}h_{*,L}.
\label{eq:h0-Lambda-h}
\end{equation}
Contracting \eqref{eq:defect-fixed-point} with \(h_{*,I}\) gives
\begin{equation}
\Lambda_h
=
48\pi^2\epsilon h_0^2 .
\label{eq:Lambda-h-fixed-point}
\end{equation}
This relation fixes the one-loop contractions in the defect-direction
sector.

\subsection{Defect-direction operators and perturbative setup}
\label{subsec:defect-direction-operators}

For the lowest members we use
\begin{equation}
\phi_0^{\rm bare}
=
h_*^I\phi_I,
\qquad
D_\mu^{\rm bare}
=
h_*^I\partial_\mu\phi_I,
\label{eq:phi0-D-bare}
\end{equation}
and
\begin{equation}
\mathcal D_{\mu\nu}^{\rm bare}
=
\frac{h_*^I}{\sqrt{2}}
\left(
\partial_\mu\partial_\nu
-
\frac{\delta_{\mu\nu}}{d-1}
\partial_\perp^2
\right)
\phi_I .
\label{eq:D2-bare}
\end{equation}
The renormalized operators are defined by
\begin{equation}
\phi_0^{\rm bare}
=
Z_{\phi_0}\phi_0,
\qquad
D_\mu^{\rm bare}
=
Z_DD_\mu,
\qquad
\mathcal D_{\mu\nu}^{\rm bare}
=
Z_{\mathcal D}\mathcal D_{\mu\nu}.
\label{eq:universal-Z-convention}
\end{equation}
In the supplemental model tables, the renormalized radial elementary
operator is normalized as \(\Phi_0=\phi_0/h_0\); hence
\(\phi_0=h_0\Phi_0\). Unit-normalized OPE coefficients are unchanged
by this rescaling, while the two-point coefficients obey
\(C_{\phi_0}=h_0^2C_{\Phi_0}\).

At the fixed point, \(D_\mu\) is the displacement operator defined by
\eqref{eq:displacement-ward-integral}.  The symmetric-traceless
\(\mathcal D_{\mu\nu}\) is the transverse-spin-two component of the second
transverse derivative. The scalar trace is related to other operators by
the bulk equation of motion.
The free propagator restricted to the defect is
\begin{align}
 \begin{tikzpicture}[baseline=-0.5ex]
  \draw[dashed] (0,0) -- (0.8,0);
 \end{tikzpicture}
 \;\, \equiv \;
 \left\langle
 \phi_I(\tau_1)\phi_J(\tau_2)
 \right\rangle_{\lambda=h=0}
 &=
 \frac{\kappa\delta_{IJ}}
 {(\tau_{12}^2)^{1-\frac{\epsilon}{2}}},
&
\kappa
&=
\frac{\Gamma\!\left(\frac d2\right)}
{2\pi^{d/2}(d-2)} .
\label{eq:free-prop}
\end{align}
The perturbative expansion contains
\begin{align}
 \begin{tikzpicture}[baseline=-0.5ex]
  \draw[dashed] ( 0.2, 0.2) -- (-0.2,-0.2);
  \draw[dashed] (-0.2, 0.2) -- ( 0.2,-0.2);
  \node at (0,0) [bcirc] {};
 \end{tikzpicture}
 \;\; &\equiv\;
 -\lambda^{IJKL}\int d^d x\,\ldots ,
&
 \begin{tikzpicture}[baseline=-0.5ex]
  \draw[thick] (-0.1,0) -- (0.7,0);
  \draw[dashed] (0.3,0) -- (0.3,0.6);
  \node at (0.3,0) [dcirc] {};
 \end{tikzpicture}
 \;\; &\equiv\;
 -h_I\int_{-\infty}^{\infty}d\tau\,\ldots .
\label{eq:universal-feynman-rules}
\end{align}

\subsection{Two- and three-point functions}
\label{subsec:universal-two-three}

For
\(\mathcal O=\phi_0,D_\mu,\mathcal D_{\mu\nu}\),
\begin{equation}
\left\langle
\mathcal O^{\rm bare}(\tau_1)
\mathcal O^{\rm bare}(\tau_2)
\right\rangle
=
h_0^2\,\phiHphiHdiagA
+
\frac{\Lambda_h}{16\pi^2}\,
\phiHphiHdiagB
+
O(\lambda^2).
\label{eq:bare-twopt-universal-form}
\end{equation}

The explicit results are
\begin{equation}
\begin{array}{c|c|c}
\text{operator}
&
\phiHphiHdiagA
&
\phiHphiHdiagB
\\
\hline
\phi_0
&
\displaystyle
\frac{\kappa}{\tau^{2-\epsilon}}
&
\displaystyle
-\frac{3}{(4\pi)^2}
\frac{1}{\tau^{2-3\epsilon}}
\left(
\frac{2}{3\epsilon}
-\frac{1}{3}
+\aleph
+O(\epsilon)
\right)
\\[3mm]
\hline
D_\mu
&
\displaystyle
\delta_{\mu\nu}(2-\epsilon)
\frac{\kappa}{\tau^{4-\epsilon}}
&
\displaystyle
-\delta_{\mu\nu}
\frac{3}{(4\pi)^2}
\frac{1}{\tau^{4-3\epsilon}}
\left(
\frac{4}{9\epsilon}
-\frac{8}{27}
+\frac{2}{3}\aleph
+O(\epsilon)
\right)
\\[3mm]
\hline
\mathcal D_{\mu\nu}
&
\displaystyle
\Pi_{\mu\nu,\rho\sigma}
(2-\epsilon)(4-\epsilon)
\frac{\kappa}{\tau^{6-\epsilon}}
&
\displaystyle
-\Pi_{\mu\nu,\rho\sigma}
\frac{3}{2(4\pi)^2}
\frac{1}{\tau^{6-3\epsilon}}
\left(
\frac{32}{15\epsilon}
-\frac{424}{225}
+\frac{16}{5}\aleph
+O(\epsilon)
\right)
\end{array}
\label{eq:defect-direction-diagram-table}
\end{equation}
where
\begin{equation}
\Pi_{\mu\nu,\rho\sigma}
=
\frac12
\left(
\delta_{\mu\rho}\delta_{\nu\sigma}
+
\delta_{\mu\sigma}\delta_{\nu\rho}
\right)
-
\frac{1}{d-1}
\delta_{\mu\nu}\delta_{\rho\sigma},
\qquad
\aleph
=
1+\gamma_{\rm E}+\log\pi .
\label{eq:spin-two-projector}
\end{equation}

The resulting CFT data are
\begin{table}[h]
\centering
\small
\renewcommand{\arraystretch}{1.7}
\begin{tabular}{c|c|c|c}
\hline
operator
&
\(Z_{\mathcal O}\)
&
\(\Delta_{\mathcal O}\)
&
\(C_{\mathcal O}\)
\\
\hline
\(\phi_0\)
&
\(\displaystyle
1-\frac{\Lambda_h}
{64\pi^2\epsilon h_0^2}
+O(\lambda^2)
\)
&
\(\displaystyle
1+\epsilon+O(\epsilon^2)
\)
&
\(\displaystyle
\frac{h_0^2}{4\pi^2}
\left[
1-\frac{\epsilon}{2}(1+2\aleph)
\right]
+O(\epsilon^2)
\)
\\[3mm]
\hline
\(D_\mu\)
&
\(\displaystyle
1-\frac{\Lambda_h}
{192\pi^2\epsilon h_0^2}
+O(\lambda^2)
\)
&
\(\displaystyle
2
\)
&
\(\displaystyle
\frac{h_0^2}{2\pi^2}
\left(
1-\frac{7\epsilon}{6}
\right)
+O(\epsilon^2)
\)
\\[3mm]
\hline
\(\mathcal D_{\mu\nu}\)
&
\(\displaystyle
1-\frac{\Lambda_h}
{320\pi^2\epsilon h_0^2}
+O(\lambda^2)
\)
&
\(\displaystyle
3-\frac{\epsilon}{5}
+O(\epsilon^2)
\)
&
\(\displaystyle
\frac{2h_0^2}{\pi^2}
\left[
1-\frac{\epsilon}{25}(34-5\aleph)
\right]
+O(\epsilon^2)
\)
\\
\hline
\end{tabular}
\caption{One-loop two-point CFT data for the lowest defect-direction
operators.}
\label{tab:defect-direction-two-point-data}
\end{table}

At the order needed here there is a single connected three-point topology,
\begin{equation}
\left\langle
\mathcal O_1^{\rm bare}(\tau_1)
\mathcal O_2^{\rm bare}(\tau_2)
\mathcal O_3^{\rm bare}(\tau_3)
\right\rangle
=
\frac{\Lambda_h}{16\pi^2}\,
\threeptatep
+
O(\lambda^2).
\label{eq:bare-3pt-diagram}
\end{equation}

The results are
\begin{table}[h]
\centering
\small
\renewcommand{\arraystretch}{1.7}
\begin{tabular}{c|c|c|c}
\hline
three-point function
&
\(\threeptatep\)
&
\(T_{ABC}\)
&
\(\lambda_{ABC}\)
\\
\hline
\(\left\langle\phi_0\phi_0\phi_0\right\rangle\)
&
\(\displaystyle
\frac{1}{8\pi^2}
\frac{1}{|\tau_{12}\tau_{13}\tau_{23}|}
\)
&
\(1\)
&
\(\displaystyle
\frac{3\pi\epsilon}{h_0}
\)
\\[3mm]
\hline
\(\left\langle\phi_0D_\mu D_\nu\right\rangle\)
&
\(\displaystyle
\frac{1}{12\pi^2}
\frac{\delta_{\mu\nu}}
{|\tau_{12}||\tau_{13}||\tau_{23}|^3}
\)
&
\(\delta_{\mu\nu}\)
&
\(\displaystyle
\frac{\pi\epsilon}{h_0}
\)
\\[3mm]
\hline
\(\left\langle
\phi_0\mathcal D_{\mu\nu}\mathcal D_{\rho\sigma}
\right\rangle\)
&
\(\displaystyle
\frac{1}{5\pi^2}
\frac{\Pi_{\mu\nu,\rho\sigma}}
{|\tau_{12}||\tau_{13}||\tau_{23}|^5}
\)
&
\(\Pi_{\mu\nu,\rho\sigma}\)
&
\(\displaystyle
\frac{3\pi\epsilon}{5h_0}
\)
\\[3mm]
\hline
\(\left\langle
D_\mu D_\nu\mathcal D_{\rho\sigma}
\right\rangle\)
&
\(\displaystyle
\frac{\sqrt{2}}{15\pi^2}
\frac{\Pi_{\mu\nu,\rho\sigma}}
{|\tau_{12}||\tau_{13}|^3|\tau_{23}|^3}
\)
&
\(\Pi_{\mu\nu,\rho\sigma}\)
&
\(\displaystyle
\frac{2\pi\epsilon}{5h_0}
\)
\\
\hline
\end{tabular}
\caption{Three-point data for the lowest defect-direction operators.}
\label{tab:universal-three-point-data}
\end{table}

\subsection{Four-point functions and consistency conditions}
\label{subsec:universal-four-point}

The four-point functions contain the three pairwise disconnected
contractions together with a connected contribution,
\begin{align}
\left\langle
\mathcal O_1\mathcal O_2\mathcal O_3\mathcal O_4
\right\rangle_{\rm bare}
&=
\left\langle
\mathcal O_1\mathcal O_2
\right\rangle_{\rm bare}
\left\langle
\mathcal O_3\mathcal O_4
\right\rangle_{\rm bare}
+
\left\langle
\mathcal O_1\mathcal O_3
\right\rangle_{\rm bare}
\left\langle
\mathcal O_2\mathcal O_4
\right\rangle_{\rm bare}
\nonumber\\
&\quad
+
\left\langle
\mathcal O_1\mathcal O_4
\right\rangle_{\rm bare}
\left\langle
\mathcal O_2\mathcal O_3
\right\rangle_{\rm bare}
+
\left\langle
\mathcal O_1\mathcal O_2\mathcal O_3\mathcal O_4
\right\rangle_{\rm conn}.
\label{eq:four-point-disconnected-connected}
\end{align}
At leading nontrivial order,
\begin{equation}
\left\langle
\mathcal O_1^{\rm bare}
\mathcal O_2^{\rm bare}
\mathcal O_3^{\rm bare}
\mathcal O_4^{\rm bare}
\right\rangle_{\rm conn}
=
\frac{\Lambda_h}{16\pi^2}\,
\connectedfourpt
+
O(\lambda^2).
\label{eq:bare-4pt-diagram}
\end{equation}

The connected integrals are
\begingroup
\setlength{\intextsep}{6pt}
\begin{table}[H]
\centering
\small
\renewcommand{\arraystretch}{1.7}
\begin{tabular}{c|c}
\hline
four-point function
&
\(\connectedfourpt\)
\\
\hline
\(\left\langle
\phi_0\phi_0\phi_0\phi_0
\right\rangle\)
&
\(\displaystyle
\frac{1}{8\pi^4}
\frac{I_1(\chi)}
{\tau_{12}^2\tau_{34}^2}
\)
\\[3mm]
\hline
\(\left\langle
\phi_0\phi_0D_\mu D_\nu
\right\rangle\)
&
\(\displaystyle
-\delta_{\mu\nu}
\frac{1}{24\pi^4}
\frac{I_2(\chi)}
{\tau_{12}^2\tau_{34}^4}
\)
\\[3mm]
\hline
\(\left\langle
\phi_0\phi_0
\mathcal D_{\mu\nu}\mathcal D_{\rho\sigma}
\right\rangle\)
&
\(\displaystyle
\Pi_{\mu\nu,\rho\sigma}
\frac{1}{30\pi^4}
\frac{I_3(\chi)}
{\tau_{12}^2\tau_{34}^6}
\)
\\[3mm]
\hline
\(\left\langle
D_\mu D_\nu D_\rho D_\sigma
\right\rangle\)
&
\(\displaystyle
\left(
\delta_{\mu\nu}\delta_{\rho\sigma}
+
\delta_{\mu\rho}\delta_{\nu\sigma}
+
\delta_{\mu\sigma}\delta_{\nu\rho}
\right)
\frac{1}{30\pi^4}
\frac{I_4(\chi)}
{\tau_{12}^4\tau_{34}^4}
\)
\\
\hline
\end{tabular}
\caption{Connected four-point integrals for the lowest defect-direction
operators.}
\label{tab:universal-four-point-integrals}
\end{table}
\endgroup

The following expressions are written for \(0<\chi<1\).
For crossing to other real ordering regions, the logarithms are understood
as \(\log|\chi|\) and \(\log|1-\chi|\), with the conformal prefactors and
tensor structures transformed simultaneously.
The functions are
\begin{align}
I_1(\chi)
&=
\chi\log(1-\chi)
+
\frac{\chi^2}{1-\chi}\log\chi ,
\nonumber\\
I_2(\chi)
&=
\frac{\chi^2}{1-\chi}
-
(\chi+2)\chi\log(1-\chi)
+
\frac{\chi^4}{(1-\chi)^2}\log\chi ,
\nonumber\\
I_3(\chi)
&=
\frac{(2\chi^2+3\chi-3)\chi^2}
{2(1-\chi)^2}
+
(\chi^2+3\chi+6)\chi\log(1-\chi)
+
\frac{\chi^6}{(1-\chi)^3}
\log\chi  ,
\nonumber\\
I_4(\chi)
&=
\frac{(\chi^2-\chi+1)\chi^2}{(1-\chi)^2}
+
\frac12
(2\chi^2+\chi+2)\chi\log(1-\chi)
+
\frac{(2\chi^2-5\chi+5)\chi^4}
{2(1-\chi)^3}
\log\chi  .
\label{eq:universal-I-functions}
\end{align}

These correlators provide an explicit realization of the crossing, OPE,
and displacement conditions.

\paragraph{Crossing symmetry.}

Substituting \eqref{eq:universal-I-functions} into the applicable crossing
relations, all tensor-component crossing residuals vanish through the
perturbative order considered.

\paragraph{OPE consistency.}

Consider transverse-scalar primaries \(X_a\) with
\(n_X=2\) in the \(\phi_0\times\phi_0\) OPE,
\begin{equation}
\Delta_{X_a}
=
2+\epsilon\gamma_a+O(\epsilon^2),
\qquad
\lambda_{\phi_0\phi_0X_a}^2
=
\left(
\lambda_{\phi_0\phi_0X_a}^{(0)}
\right)^2
+
\epsilon
\left(
\lambda_{\phi_0\phi_0X_a}^{2}
\right)^{(1)}
+
O(\epsilon^2).
\label{eq:n2-ope-data}
\end{equation}
Their contribution begins as
\begin{align}
\sum_a
\lambda_{\phi_0\phi_0X_a}^2
g_{\Delta_{X_a}}(\chi)
={}&
\chi^2
\Biggl[
\sum_a
\left(
\lambda_{\phi_0\phi_0X_a}^{(0)}
\right)^2
+
\epsilon
\sum_a
\left(
\lambda_{\phi_0\phi_0X_a}^{2}
\right)^{(1)}
\nonumber\\
&\hspace{1.7cm}
+
\epsilon\log\chi
\sum_a
\left(
\lambda_{\phi_0\phi_0X_a}^{(0)}
\right)^2
\gamma_a
\Biggr]
+
O(\chi^3,\epsilon^2).
\label{eq:n2-phi4-block-expansion}
\end{align}

The four-scalar correlator gives
\begin{equation}
\mathcal G_{\phi_0\phi_0\phi_0\phi_0}(\chi)
=
1
+
2\chi^2
+
\epsilon\chi^2
\left[
\left(
4+\frac{6}{h_0^2}
\right)\log\chi
-
\frac{6}{h_0^2}
\right]
+
O(\chi^3,\epsilon^2),
\label{eq:phi4-n2-expansion}
\end{equation}
Comparing the coefficients of \(\chi^2\),
\(\epsilon\chi^2\log\chi\), and \(\epsilon\chi^2\) with
\eqref{eq:n2-phi4-block-expansion} fixes the leading sum of squared OPE
coefficients, the anomalous-dimension-weighted sum, and the first correction
to the sum of squares, respectively.

The mixed correlator gives
\begin{equation}
\mathcal G_{\phi_0\phi_0DD}(\chi)
=
1
-
\frac{3\epsilon}{h_0^2}\chi^2
+
O(\chi^3,\epsilon^2),
\label{eq:phi2D2-n2-expansion}
\end{equation}
The coefficient of \(\epsilon\chi^2\) fixes the first-order sum of
mixed OPE products.

The scalar projection of the four-displacement correlator contains no
\(\chi^2\) term through \(O(\epsilon)\). At leading order this implies
\(\sum_a(\lambda_{DDX_a}^{(0)})^2=0\). In a real orthonormal basis,
each term is nonnegative, so
\begin{equation}
\lambda_{DDX_a}^{(0)}=0
\label{eq:DDX-leading-zero}
\end{equation}
for every exchanged dimension-two scalar.

The dimension-two scalar sector therefore obeys the following
basis-independent OPE constraints.
\begin{equation}
\begin{aligned}
\sum_a
\left(
\lambda_{\phi_0\phi_0X_a}^{(0)}
\right)^2
&=
2,
 &\qquad 
\sum_a
\left(
\lambda_{\phi_0\phi_0X_a}^{(0)}
\right)^2
\gamma_a
&=
4+\frac{6}{h_0^2},
\\
\sum_a
\left(
\lambda_{\phi_0\phi_0X_a}^{2}
\right)^{(1)}
&=
-\frac{6}{h_0^2},
 &\qquad 
\sum_a
\left(
\lambda_{\phi_0\phi_0X_a}
\lambda_{DDX_a}
\right)^{(1)}
&=
-\frac{3}{h_0^2},
\\
\lambda_{DDX_a}^{(0)}
&=
0.
\label{eq:n2-universal-ope-sum-rules}
\end{aligned}
\end{equation}
These are basis-independent constraints on the complete dimension-two scalar
sector.  In every model and every defect branch considered in this work,
their left-hand sides evaluated from the independently computed mixing
matrices and three-point coefficients reproduce the right-hand sides.

\paragraph{Integral identities.}

We apply the displacement identities to
\begin{equation}
\left\langle
D_\mu D_\nu\phi_0\phi_0
\right\rangle_c,
\qquad
\left\langle
D_\mu D_\nu D_\rho D_\sigma
\right\rangle_c .
\label{eq:universal-integral-correlators}
\end{equation}
At the orders relevant here,
\begin{equation}
D_\mu\times\phi_0
\ \supset\
D_\mu,
\qquad
D_\mu\times D_\nu
\ \supset\
\phi_0,\quad X_a,\quad\mathcal D_{\mu\nu}.
\label{eq:universal-endpoint-content}
\end{equation}

For \(X=\phi_0,\mathcal D\), the higher-order predictions use
\begin{equation}
\lambda_{DDX}=\epsilon\lambda_{DDX}^{(1)}+\epsilon^2\lambda_{DDX}^{(2)}+O(\epsilon^3),
\qquad
\Delta_X-n_X=\epsilon\gamma_X+\epsilon^2\gamma_X^{(2)}+O(\epsilon^3).
\label{eq:universal-higher-endpoint-data}
\end{equation}
For the defect fixed points with \(h_*^I\neq0\) considered here,
\(h_0\) defined in \eqref{eq:h0-Lambda-h} is \(O(1)\). Retaining \(h_0\) for its
leading-order value, we include the next-order correction through
\begin{equation}
h_0\ \longrightarrow\ h_0+\epsilon h_0^{(1)}+O(\epsilon^2).
\label{eq:universal-h0-expansion}
\end{equation}
The OPE coefficients include this correction in their expansion.

For the mixed correlator, both sides of the non-logarithmic identities
\eqref{eq:displacement-nonlog-01} vanish at \(O(1)\), so they are
satisfied automatically. At \(O(\epsilon)\), the connected correlator in
table~\ref{tab:universal-four-point-integrals} gives the constraint
\begin{equation}
-\frac{
\bigl(\lambda_{DD\phi_0}^{(1)}\bigr)^2
}{
\gamma_{\phi_0}
}
+
\frac{
\lambda_{DD\phi_0}^{(1)}
\lambda_{\phi_0\phi_0\phi_0}^{(1)}
}{
\gamma_{\phi_0}
}
=
\frac{2\pi^2}{h_0^2}.
\label{eq:mixed-nonlog-constraint}
\end{equation}
For the logarithmic identity, the connected mixed correlator also
vanishes at \(O(1)\), but the enhanced endpoint contribution survives.
Matching it to the right-hand side of
\eqref{eq:displacement-log-identities} gives
\begin{equation}
\frac{
\bigl(\lambda_{DD\phi_0}^{(1)}\bigr)^2
}{
\gamma_{\phi_0}^2
}
=
\frac{\pi^2}{h_0^2}.
\label{eq:mixed-log-constraint}
\end{equation}

At \(O(\epsilon)\), the same logarithmic identity gives the prediction
\begin{equation}
\lambda_{DD\phi_0}^{(2)}-\frac{\pi}{h_0}\gamma_{\phi_0}^{(2)}
=\frac{\pi}{12h_0}-\frac{\pi h_0^{(1)}}{h_0^2}.
\label{eq:mixed-log-higher-prediction}
\end{equation}
This predicts a combination of second-order OPE and dimension corrections,
rather than directly checking those uncomputed corrections.

For the four-displacement correlator, the non-logarithmic identities
at \(O(\epsilon)\) give
\begin{align}
2
\frac{
\bigl(\lambda_{DD\phi_0}^{(1)}\bigr)^2
}{
\gamma_{\phi_0}
}
+
\frac{
\bigl(\lambda_{DD\mathcal D}^{(1)}\bigr)^2
}{
\gamma_{\mathcal D}
}
=
\frac{6\pi^2}{5h_0^2}
,
\label{eq:D4-nonlog-constraint-1}
\\[2mm]
\frac{
\bigl(\lambda_{DD\phi_0}^{(1)}\bigr)^2
}{
\gamma_{\phi_0}
}
+
\frac{
\bigl(\lambda_{DD\mathcal D}^{(1)}\bigr)^2
}{
\gamma_{\mathcal D}
}
=
\frac{\pi^2}{5h_0^2}
.
\label{eq:D4-nonlog-constraint-2}
\end{align}
The connected four-displacement correlator also vanishes at \(O(1)\),
so the logarithmic identities at this order constrain the endpoint data
alone and give
\begin{equation}
\begin{aligned}
2
\frac{
\bigl(\lambda_{DD\phi_0}^{(1)}\bigr)^2
}{
\gamma_{\phi_0}^2
}
-
\frac{
\bigl(\lambda_{DD\mathcal D}^{(1)}\bigr)^2
}{
\gamma_{\mathcal D}^2
}
&=
-\frac{2\pi^2}{h_0^2},
\\[2mm]
\frac{
\bigl(\lambda_{DD\phi_0}^{(1)}\bigr)^2
}{
\gamma_{\phi_0}^2
}
&=
\frac{\pi^2}{h_0^2},
 &\qquad 
\frac{
\bigl(\lambda_{DD\phi_0}^{(1)}\bigr)^2
}{
\gamma_{\phi_0}^2
}
-
\frac{
\bigl(\lambda_{DD\mathcal D}^{(1)}\bigr)^2
}{
\gamma_{\mathcal D}^2
}
&=
-\frac{3\pi^2}{h_0^2}.
\label{eq:D4-log-constraints}
\end{aligned}
\end{equation}

At \(O(\epsilon)\), the same logarithmic identities reproduce
\eqref{eq:mixed-log-higher-prediction} and give the additional prediction
\begin{equation}
\lambda_{DD\mathcal D}^{(2)}+\frac{2\pi}{h_0}\gamma_{\mathcal D}^{(2)}
=\frac{23\pi}{150h_0}-\frac{2\pi h_0^{(1)}}{5h_0^2}.
\label{eq:D4-log-higher-prediction}
\end{equation}

The independently computed data are
\begin{equation}
\gamma_{\phi_0}=1,
\qquad
\gamma_{\mathcal D}=-\frac15,
\qquad
\lambda_{DD\phi_0}^{(1)}
=
\frac{\pi}{h_0},
\qquad
\lambda_{\phi_0\phi_0\phi_0}^{(1)}
=
\frac{3\pi}{h_0},
\qquad
\lambda_{DD\mathcal D}^{(1)}
=
\frac{2\pi}{5h_0}.
\label{eq:integral-independent-cft-data}
\end{equation}
They satisfy the leading-order relations above; the higher-order
logarithmic relations predict corrections not computed directly here.

The dimension-two scalar sector \(X_a\) has
\begin{equation}
\Delta_{X_a}
=
2+\epsilon\gamma_a+O(\epsilon^2),
\qquad
\lambda_{DDX_a}
=
\epsilon\lambda_{DDX_a}^{(1)}
+
O(\epsilon^2).
\label{eq:n2-integral-data}
\end{equation}
Its complete contribution to the non-log identities begins at
\(O(\epsilon^2)\). In the logarithmic combinations considered here,
the enhanced dimension-two endpoint terms cancel after all ordering
regions are combined.


\section{Model-dependent data and consistency conditions}
\label{sec:model-dependent-checks}

In this section, we present model-specific CFT data and test them using
the consistency conditions of sections~\ref{sec:consistency-conditions}
and~\ref{sec:universal-sector}.
Their operator bases, mixing matrices, and
correlation functions are given in the supplemental material. The operator
range is specified in \eqref{eq:intro-dimension-range}, with additional
\(DDX\) three-point coefficients and anomalous dimensions for the relevant
exchanges with \(n_X\leq3\), as described in
section~\ref{sec:introduction}.
The four-point functions have the identical-operator or paired-operator
content specified in section~\ref{subsec:crossing}.

The \(O(N)\) model provides a continuous-symmetry benchmark. The
hypercubic singlet sector exhibits the basis-independent OPE constraints
in a mixing block, while biconical Case I realizes tilt identities for two
broken continuous symmetry factors. The hypertetrahedral and \(MN\)
models supply further symmetry-breaking patterns, mixing sectors, and
defect branches. The explicit examples illustrate the conditions used
across the dataset.
Further details of the bulk models and the classification of their defect
fixed points can be found in~\cite{Pannell:2023pwz}.

\subsection{\texorpdfstring{\(O(N)\)}{O(N)} model}
\label{subsec:ON-model-checks}

The \(O(N)\)-invariant quartic tensor is
\begin{equation}
\lambda^{IJKL}=\lambda\Delta^{IJKL},
\qquad
\Delta^{IJKL}=\delta^{IJ}\delta^{KL}
+
\delta^{IK}\delta^{JL}
+
\delta^{IL}\delta^{JK},
\label{eq:ON-invariant-quartic-tensor}
\end{equation}
with Wilson--Fisher fixed point
\begin{equation}
\lambda_*
=
\frac{16\pi^2\epsilon}{N+8}
+
O(\epsilon^2).
\end{equation}
Using the bulk symmetry, a representative nontrivial defect fixed point can
be chosen as
\begin{equation}
h_*^I=h_0\,\delta^{IN},
\end{equation}
so that
\begin{equation}
O(N)\longrightarrow O(N-1)
\end{equation}
on the defect.  The scalar along the defect direction is a singlet, while
the \(N-1\) orthogonal elementary scalars form the tilt multiplet associated
with the broken generators.

The \(O(N)\) data in the supplemental material form a subset of the
results in~\cite{Belton:2025hbu}. We use this subset as a benchmark for our
calculations and verification procedure before applying them to the
remaining four models, whose data are new.

\subsection{Hypercubic model}
\label{subsec:hypercubic-model-checks}

The hypercubic quartic tensor may be written as
\begin{equation}
\lambda^{IJKL}
=
\lambda\,\Delta^{IJKL}
+
g\,\delta^{IJKL},
\label{eq:hypercubic-lambda-sec4}
\end{equation}
Here \(\Delta^{IJKL}\) is the tensor in
\eqref{eq:ON-invariant-quartic-tensor}, and \(\delta^{IJKL}\) is nonzero
only when all four indices coincide.
At the hypercubic bulk fixed point,
\begin{equation}
\lambda_*
=
\frac{16\pi^2\epsilon}{3N}
+
O(\epsilon^2),
\qquad
g_*
=
\frac{16\pi^2(N-4)\epsilon}{3N}
+
O(\epsilon^2).
\label{eq:hypercubic-bulk-fp-sec4}
\end{equation}

For generic \(N\ne4\), the defect fixed-point equation has
\begin{equation}
1+\sum_{p=1}^{N}
\binom{N}{p}2^p
=
3^N
\end{equation}
solutions, which fall into \(N+1\) equivalence classes under
the bulk hypercubic symmetry.
A representative with \(p\) nonzero defect-coupling components may be chosen as
\begin{equation}
h_*^I
=
\frac{h_0}{\sqrt p}
\left(
\underbrace{1,\ldots,1}_{p},
0,\ldots,0
\right),
\label{eq:hypercubic-defect-vector-sec4}
\end{equation}
with preserved symmetry
\begin{equation}
S_p\times B_{N-p}.
\end{equation}
On these generic branches the internal symmetry is discrete, so there are
no tilt operators. At \(N=4\), the leading-order anisotropic coupling
vanishes and the bulk symmetry is enhanced to \(O(4)\); the discrete
fixed-point count and absence of tilt operators do not apply to that limit.

\paragraph{Dimension-two singlet operators and OPE consistency.}

For the generic hypercubic defect with \(1<p<N\), three independent
dimension-two scalar bilinears are singlets under the preserved symmetry
and can mix. This sector provides an explicit test of OPE reconstruction
in the presence of operator mixing.
The bare representative of the defect-direction field \(\Phi_0\)
introduced in section~\ref{sec:universal-sector} is
\(\Phi_0^{\mathrm{bare}}=p^{-1/2}\sum_{i=1}^{p}\phi_i\).
We choose the bare basis operators
\begin{equation}
S_A^{\mathrm{bare}}
=
\frac{1}{\sqrt p}\sum_{i=1}^{p}\phi_i^2,
\quad
P_A^{\mathrm{bare}}
=
\frac{1}{\sqrt{p(p-1)}}
\sum_{\substack{i,j=1\\i\ne j}}^{p}\phi_i\phi_j,
\quad
S_I^{\mathrm{bare}}
=
\frac{1}{\sqrt{N-p}}
\sum_{i=p+1}^{N}\phi_i^2.
\label{eq:hypercubic-singlet-bare}
\end{equation}
All three operators are singlets under the preserved
\(S_p\times B_{N-p}\) symmetry and can therefore mix under renormalization.
This is the \((S_A,P_A,S_I)\) basis used in the supplemental hypercubic
mixing tables.

At \(p=1\), only one field has a nonzero defect coupling, so \(P_A\) is
absent and the singlet basis reduces to \((S_A,S_I)\). At \(p=N\),
all fields have nonzero defect couplings, so \(S_I\) is absent and the basis reduces
to \((S_A,P_A)\). The mixing problem at either endpoint is formulated
in this reduced basis, without using the singular normalization factors
in \eqref{eq:hypercubic-singlet-bare}.

The eigenoperators of the one-loop mixing matrix are
\begin{equation}
X_\alpha
=
(v_\alpha)_1 S_A
+
(v_\alpha)_2 P_A
+
(v_\alpha)_3 S_I,
\qquad
\alpha=1,2,3,
\label{eq:hypercubic-singlet-primaries}
\end{equation}
where the bilinears on the right-hand side denote
renormalized basis operators and \(v_\alpha\) are the
normalized eigenvectors of the mixing matrix
\begin{equation}
k_{\mathrm{sing}}
=
\begin{pmatrix}
k_{11} & k_{12} & k_{13}\\
k_{12} & k_{22} & 0\\
k_{13} & 0 & k_{33}
\end{pmatrix}.
\label{eq:hypercubic-singlet-k}
\end{equation}
Its nonvanishing entries are
\begin{equation}
\begin{aligned}
k_{11}
&=
\frac{(N+p-2)(10N+3p-4)}
{3N(N+3p-4)},
\\
k_{12}
&=
\frac{6\sqrt{p-1}}{N+3p-4},
 &\qquad 
k_{13}
&=
\frac{\sqrt{p(N-p)}}{3N},
\\
k_{22}
&=
\frac{9N^2+27Np-52N+6p-8}
{3N(N+3p-4)},
\\
k_{33}
&=
\frac{2N^2+14Np-10N-3p^2-2p+8}
{3N(N+3p-4)}.
\label{eq:hypercubic-singlet-k-elements}
\end{aligned}
\end{equation}

Let \(\kappa_\alpha\) denote the eigenvalues of
\(k_{\mathrm{sing}}\),
\begin{equation}
k_{\mathrm{sing}}v_\alpha
=
\kappa_\alpha v_\alpha,
\qquad
v_\alpha^{\mathsf T}v_\beta
=
\delta_{\alpha\beta}.
\label{eq:hypercubic-singlet-eigenvalue}
\end{equation}
For the generic branch, the normalized eigenvectors can be chosen as
\begin{equation}
v_\alpha
=
\frac{1}{\mathcal N_\alpha}
\left(
1,\,
\frac{k_{12}}{\kappa_\alpha-k_{22}},\,
\frac{k_{13}}{\kappa_\alpha-k_{33}}
\right)^{\mathsf T},
\qquad \alpha=1,2,3.
\label{eq:hypercubic-singlet-eigenvectors}
\end{equation}
Here \(\mathcal N_\alpha^2=1+k_{12}^2/(\kappa_\alpha-k_{22})^2
+k_{13}^2/(\kappa_\alpha-k_{33})^2\) fixes the normalization.
If an eigenvalue coincides with \(k_{22}\) or \(k_{33}\), this coordinate
representation is singular; a normalized eigenvector must instead be
obtained directly from \eqref{eq:hypercubic-singlet-eigenvalue}.
Substituting these eigenvectors into
\eqref{eq:hypercubic-singlet-primaries} gives the leading expressions
for the conformal primaries.
Their scaling dimensions are
\begin{equation}
\Delta_{X_\alpha}
=
2+\epsilon(\kappa_\alpha-1)
+
O(\epsilon^2).
\label{eq:hypercubic-singlet-dimensions}
\end{equation}

Using this one-loop eigenbasis and the independently computed
three-point functions, we obtain the normalized OPE
coefficients in that basis entering the
\(\phi_0\times\phi_0\) and \(D_\mu\times D_\nu\)
OPEs.
Their basis-independent combinations satisfy the universal dimension-two
scalar sum rules in
\eqref{eq:n2-universal-ope-sum-rules}.
The explicit OPE coefficients and the data for the remaining operator
sectors are collected in the supplemental material.

\subsection{Hypertetrahedral model}

The defect fixed-point classification and stability for this model are
studied in~\cite{Pannell:2024hbu}.
\label{subsec:hypertetrahedral-model-checks}

The hypertetrahedral theory has global symmetry
\begin{equation}
T_N=S_{N+1}\times\mathbb Z_2 .
\end{equation}
Let \(e_I^\alpha\), \(\alpha=1,\ldots,N+1\), denote the vertices of a regular
hypertetrahedron, normalized by
\begin{equation}
\sum_{\alpha=1}^{N+1}e_I^\alpha=0,
\qquad
\sum_{\alpha=1}^{N+1}
e_I^\alpha e_J^\alpha
=
\delta_{IJ}.
\label{eq:tetra-vertex-identities}
\end{equation}
The quartic invariant is
\begin{equation}
d^{IJKL}
=
\sum_{\alpha=1}^{N+1}
e_I^\alpha e_J^\alpha e_K^\alpha e_L^\alpha ,
\end{equation}
and
\begin{equation}
\lambda^{IJKL}
=
\lambda\,\Delta^{IJKL}
+
g\,d^{IJKL}.
\label{eq:tetra-lambda-sec4}
\end{equation}
The bulk beta functions admit two hypertetrahedral fixed points,
\begin{equation}
(\lambda_{\sigma,*},g_{\sigma,*}),
\qquad
\sigma=\pm .
\end{equation}

At leading order, the \(\sigma=\pm\) bulk fixed points are distinct
for \(N>5\) but have identical couplings at \(N=5\).
At \(N=4\), the \(\sigma=+\) branch instead reduces to the \(O(4)\)
fixed point at this order~\cite{Pannell:2023pwz}.
The discrete-symmetry description below concerns the generic
hypertetrahedral points; the enhanced-symmetry limit is treated as in the
\(O(N)\) model.

The vertex coordinates \(h^\alpha=e_I^\alpha h_*^I\) automatically
obey \(\sum_\alpha h^\alpha=0\), since
\(\sum_\alpha e_I^\alpha=0\) in \eqref{eq:tetra-vertex-identities}.
This relation expresses the redundancy of the \(N+1\) coordinates for
an \(N\)-component defect coupling.
We consider defect couplings with \(m\) equal positive entries,
\(\ell\) equal negative entries, and \(k\) zero entries, where
\begin{equation}
m+\ell+k=N+1.
\end{equation}
The zero-sum condition fixes the ratio of the positive to negative
magnitudes to \(\ell/m\), and the overall magnitude is \(h_0\)
defined in \eqref{eq:h0-Lambda-h}.
For this ansatz with two distinct nonzero coupling values and
\(g_*\neq0\), the defect fixed-point
equations require either equal nonzero block sizes or unequal nonzero
block sizes with no zero entries. We therefore discuss the following two
cases.

In Case I the two nonzero blocks have equal sizes, \(m=\ell=p\),
and the number of zero entries is \(k=N+1-2p\), which need not vanish.
Thus there are \(p\) equal positive entries and \(p\) negative entries
of the same magnitude, with \(1\leq p\leq\lfloor(N+1)/2\rfloor\).

In Case II the two nonzero blocks have unequal sizes, \(m\neq\ell\).
If \(k>0\), the beta functions of the zero components do not vanish,
so this configuration is not a defect fixed point.
Thus \(k=0\), \(m+\ell=N+1\), and the preserved permutation symmetry is
\begin{equation}
S_m\times S_\ell.
\end{equation}
For each case, the defect fixed-point equation fixes \(h_0\). The Case II family includes the stable \(N=7\), \(m=3\),
\(\ell=5\) examples in~\cite{Pannell:2023pwz}.
The supplemental data cover these two cases. The Case II operator tables
assume \(m,\ell\geq3\); the rank-two symmetric family in either block
is absent when that block has size three.

For Case I and the supplied Case II branch, at both bulk fixed points
\(\sigma=\pm\), the computed four-point functions satisfy crossing and
reproduce the basis-independent OPE combinations obtained from the
two- and three-point data. At generic hypertetrahedral points the internal symmetry is discrete,
so the integral conditions involve displacements without tilt operators.
The displacement identities are satisfied at the orders considered.

\subsection{Biconical model}
\label{subsec:biconical-model-checks}

In this subsection, hats and checks on fields and tilt operators label the
two internal-symmetry factors; the OPE coefficients retain the
unit-normalized convention defined above.

The biconical theory contains an \(m\)-component scalar
\(\widehat\phi_i\) and an \(n\)-component scalar
\(\check\phi_a\), transforming under
\begin{equation}
O(m)\times O(n).
\end{equation}
In the quartic-tensor convention of \eqref{eq:bulk-action}, its interaction
is specified by the nonzero components
\begin{equation}
\lambda^{ijkl}=\lambda_1\Delta^{ijkl},
\qquad
\lambda^{abcd}=\lambda_2\Delta^{abcd},
\qquad
\lambda^{ijab}=g\delta^{ij}\delta^{ab},
\label{eq:biconical-potential-sec4}
\end{equation}
and their index permutations. Here \(i,j,k,l\) belong to the \(O(m)\)
multiplet and \(a,b,c,d\) to the \(O(n)\) multiplet; each \(\Delta\)
is the invariant tensor in \eqref{eq:ON-invariant-quartic-tensor} within
the corresponding multiplet. 

A linear defect can couple to both scalar multiplets. We denote the
magnitudes of the fixed-point defect couplings by \(\hat{h}\) and
\(\check{h}\).
The three inequivalent branches are
\begin{align}
\text{Case I:}\qquad&
\hat{h}\neq0,
\qquad
\check{h}\neq0,
&
O(m)\times O(n)
&\longrightarrow
O(m-1)\times O(n-1),
\label{eq:biconical-case-I-sec4}
\\
\text{Case II:}\qquad&
\hat{h}\neq0,
\qquad
\check{h}=0,
&
O(m)\times O(n)
&\longrightarrow
O(m-1)\times O(n),
\label{eq:biconical-case-II-sec4}
\\
\text{Case III:}\qquad&
\hat{h}=0,
\qquad
\check{h}\neq0,
&
O(m)\times O(n)
&\longrightarrow
O(m)\times O(n-1).
\label{eq:biconical-case-III-sec4}
\end{align}

For Case I, where both defect couplings are nonzero, the defect beta
function gives
\begin{equation}
\lambda_1\hat{h}^2+g\check{h}^2=16\pi^2\epsilon,
\qquad
g\hat{h}^2+\lambda_2\check{h}^2=16\pi^2\epsilon.
\label{eq:biconical-caseI-fixed-point}
\end{equation}
For Cases II and III they reduce respectively to
\begin{equation}
\lambda_1\hat{h}^2=16\pi^2\epsilon,
\qquad
\lambda_2\check{h}^2=16\pi^2\epsilon.
\end{equation}
In Cases II and III only one of the two defect couplings is nonzero,
so their symmetry-breaking patterns are analogous to the \(O(N)\) line defect.
We therefore focus below on Case I.

In Case I both symmetry factors are broken. Continuous broken generators
are present in the first factor for \(m>1\) and in the second for \(n>1\). Choose the defect coupling in each
multiplet to point along its last component. The bare tilt
operators are
\begin{equation}
\widehat t_i^{\mathrm{bare}}=\hat h\,\widehat\phi_i,
\qquad i=1,\ldots,m-1,
\qquad
\check t_a^{\mathrm{bare}}=\check h\,\check\phi_a,
\qquad a=1,\ldots,n-1.
\label{eq:biconical-bare-tilts}
\end{equation}
Renormalized defect operators are denoted without the bare
superscript. The tilt multiplets transform as
\begin{equation}
(\mathbf{m-1},\mathbf 1),
\qquad
(\mathbf 1,\mathbf{n-1})
\end{equation}
under \(O(m-1)\times O(n-1)\).  Their two-point coefficients are
\begin{equation}
C_{\widehat t}
=
\frac{\hat{h}^2}{4\pi^2}
\left(1-\frac{\epsilon}{2}\right)
+O(\epsilon^2),
\qquad
C_{\check t}
=
\frac{\check{h}^2}{4\pi^2}
\left(1-\frac{\epsilon}{2}\right)
+O(\epsilon^2).
\label{eq:biconical-caseI-tilt-two-point}
\end{equation}

\paragraph{Four-tilt consistency conditions.}

The five conditions below are independent four-tilt projections for generic
\(m,n\geq3\). If \(m=1\), the \(\widehat t\) multiplet is absent;
if \(n=1\), the \(\check t\) multiplet is absent, and correlators requiring
that multiplet are omitted. At \(m=2\), there is a single \(\widehat t\)
component, so the corresponding broken-generator commutator and
antisymmetric tensor structure vanish; the logarithmic condition
\eqref{eq:biconical-hat-log} is then not an independent projection of the
four-tilt identity. The same applies to
\eqref{eq:biconical-check-log} at \(n=2\). The displayed scalar expressions
can still hold as algebraic relations among the radial data in these limits.

Consider
\begin{equation}
\left\langle
\widehat t_i\widehat t_j\widehat t_k\widehat t_l
\right\rangle,
\qquad
\left\langle
\check t_a\check t_b\check t_c\check t_d
\right\rangle,
\qquad
\left\langle
\widehat t_i\widehat t_j\check t_a\check t_b
\right\rangle.
\label{eq:biconical-caseI-four-tilt}
\end{equation}
Using \(I_1(\chi)\) from \eqref{eq:universal-I-functions}, the reduced correlators are
\begin{align}
\mathcal G_{\widehat t_i\widehat t_j\widehat t_k\widehat t_l}(\chi)
={}&
\delta_{ij}\delta_{kl}+\chi^2\delta_{ik}\delta_{jl}+\frac{\chi^2}{(1-\chi)^2}\delta_{il}\delta_{jk}
\nonumber\\
&\quad
+\frac{\lambda_1}{8\pi^2} I_1(\chi)(\delta_{ij}\delta_{kl}+\delta_{ik}\delta_{jl}+\delta_{il}\delta_{jk})
+O(\epsilon^2),
\label{eq:biconical-hat-four-tilt}
\\
\mathcal G_{\check t_a\check t_b\check t_c\check t_d}(\chi)
={}&
\delta_{ab}\delta_{cd}+\chi^2\delta_{ac}\delta_{bd}
+\frac{\chi^2}{(1-\chi)^2}\delta_{ad}\delta_{bc}
\nonumber\\
&\quad
+\frac{\lambda_2}{8\pi^2} I_1(\chi)
(\delta_{ab}\delta_{cd}+\delta_{ac}\delta_{bd}+\delta_{ad}\delta_{bc})
+O(\epsilon^2),
\label{eq:biconical-check-four-tilt}
\\
\mathcal G_{\widehat t_i\widehat t_j\check t_a\check t_b}(\chi)
={}&
\delta_{ij}\delta_{ab}
\left[
1+\frac{g}{8\pi^2}I_1(\chi)
\right]
+O(\epsilon^2).
\label{eq:biconical-mixed-four-tilt}
\end{align}

The relevant endpoint singularities are saturated by two scalar primaries,
\(\Phi_0\) and \(\Psi\), constructed from the defect-coupled components
of the two multiplets.
The norm in \eqref{eq:h0-Lambda-h} specializes to
\(h_0^2=\hat h^2+\check h^2\). Their bare representatives are
\begin{equation}
\Phi_0^{\mathrm{bare}}
=\frac{\hat h\,\widehat\phi_m+\check h\,\check\phi_n}{h_0},
\qquad
\Psi^{\mathrm{bare}}
=\frac{-\check h\,\widehat\phi_m+\hat h\,\check\phi_n}{h_0},
\label{eq:biconical-bare-radial-fields}
\end{equation}
Thus \(\Phi_0\) is parallel to
the defect-coupling vector and \(\Psi\) is orthogonal to it.
The radial operator of the universal sector is
\begin{equation}
\phi_0=h_0\Phi_0.
\label{eq:biconical-universal-radial-normalization}
\end{equation}
For \(\Phi_0\), we use the universal
result \(\gamma_{\Phi_0}=1\) from section~\ref{sec:universal-sector}, so
\begin{equation}
\Delta_{\Phi_0}
=
1+\epsilon+O(\epsilon^2),
\qquad
\Delta_\Psi
=
1+\epsilon\gamma_\Psi+O(\epsilon^2),
\label{eq:biconical-radial-dimensions}
\end{equation}
with
\begin{equation}
\gamma_\Psi
=
\frac{(g-\lambda_1)(g-\lambda_2)}
{\lambda_1\lambda_2-g^2}.
\label{eq:biconical-kappa}
\end{equation}
The integral constraints below concern the nondegenerate Case I
branch, with \(\hat h\check h\ne0\),
\(\lambda_1\lambda_2-g^2\ne0\), and \(\gamma_\Psi\ne0\).
At degenerate points the endpoint contributions must be treated before
expanding in \(\epsilon\).
When \(\lambda_1=\lambda_2=g\), the quartic tensor becomes \(\lambda^{IJKL}=\lambda_1\Delta^{IJKL}\),
and the internal
symmetry is enhanced to \(O(m+n)\). The orthogonal scalar \(\Psi\)
then lies in the tilt multiplet and has protected dimension one.
Equation~\eqref{eq:biconical-kappa} takes the indeterminate form \(0/0\)
at this point; the enhanced-symmetry theory must be treated directly
rather than by substituting into the generic integral constraints.
The required scalar OPE coefficients are
\begin{align}
\lambda_{\widehat t\widehat t\Phi_0}
&=\frac{\lambda_1\hat h^2+g\check h^2}{16\pi h_0},
&
\lambda_{\widehat t\widehat t\Psi}
&=\frac{(g-\lambda_1)\hat h\check h}{16\pi h_0},
\label{eq:biconical-hat-radial-ope}
\\
\lambda_{\check t\check t\Phi_0}
&=\frac{g\hat h^2+\lambda_2\check h^2}{16\pi h_0},
&
\lambda_{\check t\check t\Psi}
&=\frac{(\lambda_2-g)\hat h\check h}{16\pi h_0}.
\label{eq:biconical-check-radial-ope}
\end{align}

Using \(\gamma_{\Phi_0}=1\) from
\eqref{eq:biconical-radial-dimensions}, the three non-logarithmic
constraints are
\begin{align}
-\frac{\lambda_1}{16}
+
\frac{
\lambda_{\widehat t\widehat t\Phi_0}^{\,2}
}{\epsilon\gamma_{\Phi_0}}
+
\frac{
\lambda_{\widehat t\widehat t\Psi}^{\,2}
}{\epsilon\gamma_{\Psi}}
&=0,
\label{eq:biconical-hat-nonlog}
\\
-\frac{\lambda_2}{16}
+
\frac{
\lambda_{\check t\check t\Phi_0}^{\,2}
}{\epsilon\gamma_{\Phi_0}}
+
\frac{
\lambda_{\check t\check t\Psi}^{\,2}
}{\epsilon\gamma_{\Psi}}
&=0,
\label{eq:biconical-check-nonlog}
\\
-\frac{g}{16}
+
\frac{
\lambda_{\widehat t\widehat t\Phi_0}
\lambda_{\check t\check t\Phi_0}
}{\epsilon\gamma_{\Phi_0}}
+
\frac{
\lambda_{\widehat t\widehat t\Psi}
\lambda_{\check t\check t\Psi}
}{\epsilon\gamma_{\Psi}}
&=0.
\label{eq:biconical-mixed-nonlog}
\end{align}
The logarithmic constraints are
\begin{align}
\frac{
\lambda_{\widehat t\widehat t\Phi_0}^{\,2}
}{\epsilon^2\gamma_{\Phi_0}^2}
+
\frac{
\lambda_{\widehat t\widehat t\Psi}^{\,2}
}{\epsilon^2\gamma_{\Psi}^2}
&=
\frac{\pi^2}{\hat{h}^2},
\label{eq:biconical-hat-log}
\\
\frac{
\lambda_{\check t\check t\Phi_0}^{\,2}
}{\epsilon^2\gamma_{\Phi_0}^2}
+
\frac{
\lambda_{\check t\check t\Psi}^{\,2}
}{\epsilon^2\gamma_{\Psi}^2}
&=
\frac{\pi^2}{\check{h}^2}.
\label{eq:biconical-check-log}
\end{align}

Substituting the scalar OPE coefficients in
\eqref{eq:biconical-hat-radial-ope} and
\eqref{eq:biconical-check-radial-ope}, together with the dimensions in
\eqref{eq:biconical-radial-dimensions}, into the five constraints above
and using the Case I fixed-point equations
\eqref{eq:biconical-caseI-fixed-point} verifies all three
non-logarithmic and both logarithmic identities.

The supplemental material contains the Case I, II, and III operator data
and correlators entering the crossing, OPE, displacement, and tilt
conditions.

\subsection{\texorpdfstring{\(MN\)}{MN} model}
\label{subsec:MN-model-checks}

The \(MN\) model contains \(N=mn\) real scalar fields, arranged into
\(n\) vectors \(\vec{\phi}_r\), each with \(m\) components.
We call each vector a copy and label it by \(r=1,\ldots,n\).
The bulk symmetry is \(O(m)^n\rtimes S_n\): each \(O(m)\) factor
rotates one vector, while \(S_n\) permutes the vectors.
Writing the combined field index as \(I=(r,a)\), the quartic tensor is
\begin{equation}
\lambda^{IJKL}=\lambda\Delta^{IJKL}+g\,d_{MN}^{IJKL},
\qquad
 d_{MN}^{(r,a)(s,b)(t,c)(u,d)}
 =\delta_{rs}\delta_{rt}\delta_{ru}
 (\delta_{ab}\delta_{cd}+\delta_{ac}\delta_{bd}+\delta_{ad}\delta_{bc}),
\label{eq:MN-quartic-tensor}
\end{equation}
where \(\Delta^{IJKL}\) is defined in
\eqref{eq:ON-invariant-quartic-tensor}. Thus the quartic interaction in
\eqref{eq:bulk-action} is
\begin{equation}
\frac{1}{24}\lambda^{IJKL}\phi_I\phi_J\phi_K\phi_L
=\frac{\lambda}{8}\left(\sum_{r=1}^n\vec{\phi}_r^2\right)^2
+\frac{g}{8}\sum_{r=1}^n\left(\vec{\phi}_r^2\right)^2.
\label{eq:MN-quartic-interaction}
\end{equation}
The supplemental notebook uses loop-normalized couplings \(u,v\).
Their relation to the physical tensor couplings used here is
\begin{equation}
\lambda=16\pi^2v,
\qquad
g=16\pi^2(u-v).
\label{eq:MN-coupling-dictionary}
\end{equation}
The \(g\) term acts within each copy, whereas the \(\lambda\) term
also couples different copies.

Each vector has a defect coupling \(\vec{h}_r\) through
\eqref{eq:defect-action}. In Case I, \(\vec{h}_r\neq0\) for all
\(n\) copies. In Case II, the defect couplings are nonzero in only
\(p\) copies, with \(1\leq p<n\), and vanish in the remaining
\(n-p\) copies. Different values of \(p\) label different defect
branches~\cite{Pannell:2023pwz}.
A nonzero \(\vec{h}_r\) breaks the corresponding \(O(m)\)
factor to \(O(m-1)\), while a vanishing coupling leaves that factor
unbroken. The defect operators are organized into representations of
the symmetry preserved on each branch.
Further details can be found in the supplemental material.

\section{Discussion}
\label{sec:discussion}

We have computed defect CFT data through \(O(\epsilon)\) for the
\(O(N)\), hypercubic, hypertetrahedral, biconical, and \(MN\) models.
The datasets for the latter four models are new, while the \(O(N)\)
model provides a benchmark against existing results. We have also
identified a new universal defect-direction sector whose one-loop data
follow from the defect fixed-point equation. All applicable crossing,
OPE, and integral consistency conditions are satisfied by the computed
data at the perturbative orders considered.

Although the main contribution of this paper is technical, the three
consistency conditions are not specific to the scalar models studied
here. Crossing symmetry and OPE reconstruction apply generally to
perturbative defect CFT data, while the broken-symmetry integral
identities apply whenever the required tilt or displacement insertions
are present. Together they provide a framework for testing independently
computed defect observables across theories. The integral identities
evaluated here concern the case without anomalies. In the presence of
anomalies, further integral identities relate defect correlators to
anomaly contributions, as discussed in~\cite{Drukker:2025dfm}.

The universal sector provides a prediction from the defect fixed-point
equation: its one-loop data have the same functional dependence on \(h_0\)
across the models and branches considered. Other operator sectors retain
dependence on the bulk symmetry and the defect branch, including through
their mixing matrices and tilt multiplets.
At higher orders, diagrams involve additional contractions of bulk
coupling tensors projected along the defect direction. Such structures
already occur in the two-loop defect beta function~\cite{Pannell:2023pwz}
and are not all fixed by the leading-order defect fixed-point relation.
Further information about the bulk theory can therefore enter the
higher-order data, so the same dependence on \(h_0\) alone is not
expected to persist in general. Which relations remain universal beyond
\(O(\epsilon)\) is an open question.

Passing a consistency test confirms agreement only for the combinations
of CFT data that contribute at the order tested. It does not verify data
that first contribute at a higher order. For example, if an OPE coefficient
starts at \(O(\epsilon)\), its square enters the ordinary four-point
function only at \(O(\epsilon^2)\), so a four-point calculation through
\(O(\epsilon)\) cannot test this squared contribution.

Integral identities can nevertheless test these contributions at lower
orders: dividing an \(O(\epsilon^2)\) OPE product by an \(O(\epsilon)\)
anomalous dimension gives an \(O(\epsilon)\) term, while division by its
square gives an \(O(1)\) term in a logarithmic identity. They can also
constrain data not computed directly, as illustrated
in~\cite{Belton:2025hbu,Bliard:2023zpe,Sakkas:2024dvm}.
Section~\ref{subsec:universal-four-point} applies this predictive use to
the mixed and four-displacement correlators. Inserting the available
\(O(\epsilon)\) four-point functions into the logarithmic identities
probes the \(O(\epsilon^3)\) terms in squared OPE coefficients together
with the \(O(\epsilon^2)\) anomalous dimensions, yielding
\eqref{eq:mixed-log-higher-prediction} and
\eqref{eq:D4-log-higher-prediction}. These predict combinations of
second-order corrections; separating the OPE and dimension corrections
requires additional information. In mixed sectors the constraints
initially concern weighted sums over exchanged operators.

An open direction is to extend this analysis to line defects in
scalar--fermion theories.
Defect fixed points in the Gross--Neveu--Yukawa,
Nambu--Jona-Lasinio--Yukawa, and chiral Heisenberg models, with the same
linear coupling to bulk scalar fields, were studied in~\cite{Pannell:2023pwz}.
Yukawa interactions
introduce additional contributions to the defect beta function and
correlation functions. Determining whether the defect-direction sector
retains a common structure, and which additional bulk data enter it,
would test the scope of the universality found here.

\section*{Acknowledgments}
\addcontentsline{toc}{section}{Acknowledgments}

I am grateful to Nadav Drukker for inspiring this work and, in particular,
for suggesting the use of AI assistance in this project. I also thank him
for helpful discussions.
AI tools from OpenAI and Anthropic were used to assist with symbolic
calculations and manuscript preparation.
This work was supported in part by the INFN grant Gauge and String Theory
(GAST).

\phantomsection
\addcontentsline{toc}{section}{References}
\bibliographystyle{defect_cft_references}
\begingroup
\small
\bibliography{defect_cft_references}
\endgroup

\end{document}